\documentclass[10pt, draftcls, onecolumn, journal]{IEEEtran}
\usepackage{cite, graphicx,amssymb,color, slashbox, pict2e, multirow, rotating}
\usepackage{graphicx,amssymb}
\usepackage[tbtags]{amsmath}
\usepackage{times,amsmath}
\usepackage{subfig}
\usepackage{flushend}
\usepackage{amsmath}
\usepackage{epsfig}
\usepackage{amssymb}
\usepackage{hyperref}
\usepackage{color}
\usepackage{psfrag}
\usepackage{caption}
\usepackage{subfig}
\usepackage[usenames,dvipsnames]{xcolor}
\usepackage{textcomp}

\begin{document}
\vspace{-5mm}
\title{Sum-Rate Maximization with Minimum Power Consumption for MIMO DF Two-Way Relaying: Part II - Network Optimization}

\author{\IEEEauthorblockN{Jie Gao,
Sergiy A. Vorobyov,
Hai Jiang,
Jianshu Zhang,
and Martin Haardt
\vspace{-8mm}}
\thanks{J. Gao, S. A. Vorobyov, and H. Jiang are with the Department of
Electrical and Computer Engineering, University of Alberta,
Edmonton, AB, T6G 2V4 Canada; e-mails: {\tt\{jgao3, svorobyo,
hai1\}}@ualberta.ca. J. Zhang and M. Haardt are with the
Communication Research Laboratory, Ilmenau University of
Technology, Ilmenau, 98693, Germany; e-mails: {\tt\{Jianshu.zhang,
martin.haardt\}}@tu-ilmenau.de.

%This work has been supported in parts by the Natural Science and
%Engineering Research Council (NSERC) of Canada, the Carl Zeiss
%Award, Germany, and the Graduate School on Mobile Communications
%(GSMobicom), Ilmenau University of Technology, which is partly
%funded by the Deutsche Forschungsgemeinschaft (DFG). A part of
%this work has been performed in the framework of the European
%research project SAPHYRE, which is partly funded by the European
%Union under its FP7 ICT Objective 1.1 - The Network of the Future.
Some preliminary results of this paper were presented at
GLOBECOM~2012, Anaheim, CA, USA.}}

% make the title area
\maketitle

\begin{abstract}

In Part~II of this two-part paper, a sum-rate-maximizing power allocation with minimum power consumption is found for multiple-input multiple-output (MIMO) decode-and-forward (DF) two-way relaying (TWR) in a network optimization scenario. In this scenario, the relay and the source nodes jointly optimize their power allocation strategies to achieve network optimality. Unlike the relay optimization scenario considered in part I which features low complexity but does not achieve network optimality, the network-level optimal power allocation can be achieved in the network optimization scenario at the cost of higher complexity. The network optimization problem is considered in two cases each with several subcases. It is shown that the considered problem, which is originally nonconvex, can be transferred into different convex problems for all but two subcases. For the remaining two subcases, one for each case, it is proved that the optimal strategies for the source nodes and the relay must satisfy certain properties. Based on these properties, an algorithm is proposed for finding the optimal solution. The effect of asymmetry in the number of antennas, power limits, and channel statistics is also considered. Such asymmetry is shown to have a negative effect on both the achievable sum-rate and the power allocation efficiency in MIMO DF TWR. Simulation results demonstrate the performance of the proposed algorithm and the effect of asymmetry in the system.
\end{abstract}

\IEEEpeerreviewmaketitle

\vspace{-2mm}
\section{Introduction}

Two-way relaying (TWR) is a promising protocol featuring high spectral efficiency \cite{TWR_PRT}. Optimizing transmit strategies such as power allocation of the participating nodes in a TWR helps to maximize the spectral efficiency in terms of sum-rate \cite{PartI}-\cite{SRJoint4}. As shown in Part~I of this two-part paper \cite{PartI}, achieving the maximum sum-rate in TWR, however, does not necessarily demand the consumption of all the available power at all participating nodes. \textcolor{black}{As a result, it is of interest to find the power allocation which minimizes the power consumption of the participating nodes among all power allocations that achieve the maximum sum-rate in TWR. For brevity, this objective of optimizing the power allocation at the participating nodes is called the sum-rate maximization with minimum power consumption.} In Part I of this two-part paper, the problem of relay optimization for multiple-input multiple-output (MIMO) decode-and-forward (DF) TWR is investigated, in which the relay optimizes its own power allocation to achieve sum-rate maximization with minimum power consumption given the power allocation of the source nodes. The solution of the relay optimization problem derived in Part~I gives the optimal power allocation of the relay in a MIMO DF TWR system in the case when there is no coordination between the relay and the source nodes. Although this power allocation is in general sub-optimal on the network level, it is a viable and preferable solution for power allocation when the considered MIMO DF TWR system has limitation on the computational capability of finding the power allocation strategy. If the participating nodes have sufficient computational capability, a better performance than that in the relay optimization scenario can be achieved. In such a case, the relay and the source nodes can jointly optimize their power allocation strategies for sum-rate maximization with minimum power consumption.

Joint optimization of transmit strategies of the relay and source nodes for MIMO TWR has been studied in \cite{SRJoint1}-\cite{SRJoint4}. Transmit strategies for maximizing the weighted sum-rate of a TWR system are studied in \cite{SRJoint1}, in which the optimal solution is found through alternative optimization over the transmit strategies of the relay and source nodes. In \cite{SRJoint2}, a low-complexity sub-optimal design of relay and source node transmit strategies is derived for either sum-rate maximization or power consumption minimization under quality-of-service requirements. The joint source node and relay precoding design for minimizing the mean-square-error in a MIMO TWR system is studied in \cite{SRJoint3}. The optimal solution is found through an alternative optimization of several sub-problems obtained from the original non-convex problem. The authors in \cite{SRJoint4} solve the robust joint source and relay optimization problem for a MIMO TWR system with imperfect channel state information. Deriving the optimal solution for the joint optimization problem generally requires alternative optimization over the transmit strategies of the relay and the source nodes, which leads to high complexity \cite{SRJoint1}, \cite{SRJoint3}, \cite{SRJoint4}. All the above works consider MIMO amplify-and-forward (AF) TWR.

Considering the fact that DF TWR may achieve better performance than AF TWR, especially at low signal-to-noise ratio (SNR) \cite{ProtComp}, and the fact that DF TWR has the flexibility of performing separate power allocation/precoding for relaying the communication on each direction, it is of interest to study the problem of joint optimization over the power allocation strategies of the relay and the source nodes for MIMO DF TWR. If we further consider the power efficiency, the problem becomes more complicated. Part II of this two-part paper studies the problem of sum-rate maximization with minimum power consumption for MIMO DF TWR when the relay and the source nodes jointly optimize their power allocations. This scenario is referred to as {\it network optimization scenario}. The objective of this part is to find the joint optimal power allocation of the relay and the source nodes while reducing the complexity of finding the optimal solution. The contributions of Part~II are as follows.

First, we show that the considered network optimization problem is nonconvex. \textcolor{black}{Based on the comparison of the maximum achievable sum-rates of the multiple-access (MA) and broadcasting (BC) phases, the network optimization problem is considered for the case that the maximum achievable sum-rate of the MA phase is lager than or equal to that of the BC phase and the case that the maximum achievable sum-rate of the MA phase is less than that of the BC phase, respectively.} \textcolor{black}{In each case, we show that the original problem can be transferred, under certain conditions, into equivalent convex problem(s) which can be solved with low complexity.} \textcolor{black}{Accordingly, the above two cases are further analyzed in terms of subcases. For the subcases in which the original problem can be transferred into equivalent convex problems, the problem of sum-rate maximization and the problem of power consumption minimization are decoupled so that the sum-rate in one of the MA or BC phase is maximized while the power consumption in the other phase is minimized. The complexity of finding the optimal solution of the network optimization problem in the above subcases is therefore low.
}

Second, for the remaining two subcases in which the original problem cannot be transferred to a convex form, we prove properties that the optimal solution must satisfy. Based on these properties, we propose algorithms for finding the joint optimal power allocations for the relay and the source nodes. While the proposed algorithms find the optimal solution in iterations, the optimization problems that the rely and source nodes need to solve in each iteration are convex and simple. As a result, the complexity of the proposed algorithms for finding the optimal solution of the nonconvex joint optimization problem is acceptable in these two subcases.

Third, we demonstrate the effect of asymmetry on MIMO DF TWR in the network optimization scenario. Similar to the relay optimization scenario, we show that asymmetry in power limits, number of antennas, and channel statistics can lead to performance degradation in both the achievable sum-rate and the power allocation efficiency. Specifically, we show that the optimal power allocation in both of the aforementioned two subcases in which the original problem cannot be transferred to a convex problem is not as efficient as that in other subcases. Then, it is shown through analysis and simulation that the asymmetry in the power limits, number of antennas, and channel statistics leads to a larger occurrence probability of the above-mentioned two subcases. As a result, we show that asymmetry leads to performance degradation in the MIMO DF TWR system.

The rest of the paper is organized as follows.
Section~\ref{s:sysm} gives the system model of this work. The
network optimization problem is  studied in Section~\ref{s:netopt}. Simulation results are shown in
Section~\ref{s:simula}, and Section~\ref{s:conclu} concludes the
paper. Section~\ref{s:appen} ``Appendix'' provides proofs for the lemmas and theorems.

\section{System Model}\label{s:sysm}

A TWR with two source nodes and one relay is considered, where source
node $i\,(i=1,2)$ and the relay have $n_i$ and $n_{\rm{r}}$ antennas, respectively. The information symbol vector and the precoding matrix of source node $i$ are denoted as $\mathbf{s}_i$ and $\mathbf{W}_i$, respectively, \textcolor{black}{where $\mathbf{s}_i$ is a complex Gaussian vector with $E\{\mathbf{s}_i\}=\mathbf{0}$, $E\{\mathbf{s}_i\mathbf{s}_i^{\rm H}\}=\mathbf{I}$, and
$E\{\mathbf{s}_i\mathbf{s}_j^{\rm H}\}=\mathbf{0}$ in which the superscript $(\cdot)^{\rm H}$ stands for the conjugate transpose and $\mathbf{I}$ denotes the identity matrix.}\footnote{It is assumed as default throughout the paper that the user index $i$ and $j$ satisfy $i\neq j$.} The channels from source node $i$ to the relay and from the relay to source node $i$ are denoted as
$\mathbf{H}_{i\rm{r}}$ and $\mathbf{H}_{{\rm{r}} i}$, respectively. It is assumed that source node $i$ knows $\mathbf{H}_{{\rm{r}} i}$ and the relay knows $\mathbf{H}_{i\rm{r}}, \forall i$. It is also assumed that the relay knows $\mathbf{H}_{{\rm{r}} i}, \forall i$ by using either channel reciprocity or channel feedback.
\textcolor{black}
{For example, if the system works in the time-division duplex mode, $\mathbf{H}_{{\rm{r}} i}, \forall i$ are known at the relay due to channel reciprocity. Otherwise, when the system works in the frequency-division duplex mode, the relay needs feedback from the source nodes to obtain $\mathbf{H}_{{\rm{r}} i}, \forall i$.}

In the MA phase, source node $i$ transmits the signal $\mathbf{W}_i\mathbf{s}_i$ to the relay.  The sum-rate of the MA phase, denoted as $R^{\rm ma}(\mathbf{D})$, is bounded by \cite{Mimo_Cap}
\begin{eqnarray}\label{e:eqmacR}
\hspace{3mm}R^{\rm ma}(\mathbf{D}) %\nonumber\\
= \log{\bigg|\mathbf{I}\!+\!(\mathbf{H}_{1{\rm{r}}}
\mathbf{D}_1 \mathbf{H}_{1{\rm{r}}}^{\rm H}\!+\!
\mathbf{H}_{2{\rm{r}}} \mathbf{D}_2 \mathbf{H}_{2{\rm{r}}}^{\rm
H}) (\sigma_{{\rm{r}}}^2)^{-1}\bigg|}
\end{eqnarray}
where $\mathbf{D}_i = \mathbf{W}_i \mathbf{W}_i^{\rm H}, \forall i$, $\mathbf{D}
= [\mathbf{D}_1, \mathbf{D}_2]$ and $\sigma_{{\rm{r}}}^2\mathbf{I}$ is the noise covariance matrix at the relay.

The relay decodes $\mathbf{s}_1$ and $\mathbf{s}_2$ from the received signal, performs precoding for each of them, and then forwards the superposition of the precoded information symbols to the source nodes in the BC phase.
\textcolor{black}{
Note that the Exclusive-OR (XOR) based network coding is adopted at the relay in some works (for example \cite{XORCmp1}). While XOR based network coding may achieve better performance in terms of sum-rate than the symbol-level superposition, it relies largely on the symmetry of the traffic from the two source nodes. The asymmetry in the traffic in the two directions can lead to significant degradation in the performance of XOR in TWR \cite{XORCmp2}, \cite{XORCmp3}. As the general case of TWR is considered and there is no guarantee of traffic symmetry, the simple approach of symbol-level superposition is assumed here at the relay as it is considered in \cite{TWR_PRT}.
}

With the receiver side channel knowledge and the knowledge of the relay precoder, each source node is able to subtract its self-interference from the received signal. Denote $\mathbf{T}_{{\rm{r}}i}$ as the relay precoding matrix for relaying the signal from source node $j$ to source node $i$. Let $\mathbf{B}_i = \mathbf{T}_{{\rm{r}}i} \mathbf{T}_{{\rm{r}}i}^{\rm H}, \forall i$ and $\mathbf{B} = [\mathbf{B}_1, \mathbf{B}_2]$. Then the information rate for the communication from the relay to source node $i$, denoted as $\hat{R}_{{\rm{r}}i}(\mathbf{B}_{i})$, is expressed as
\begin{eqnarray}
\hat{R}_{{\rm{r}}i}(\mathbf{B}_{i}) = \log|\mathbf{I} +
(\mathbf{H}_{{\rm{r}}i} \mathbf{B}_{i} \mathbf{H}_{{\rm{r}}i}^{\rm
H})(\sigma_{i}^2)^{-1}|
\end{eqnarray}
where $\sigma_{i}^2\mathbf{I}$ is the noise covariance matrix at source node $i$. The sum-rate of the BC phase, denoted as $R^{\rm bc}(\mathbf{B})$, is
\begin{eqnarray}
R^{\rm bc}(\mathbf{B})=\hat{R}_{{\rm{r}}1} (\mathbf{B}_{1}) +
\hat{R}_{{\rm{r}}2} (\mathbf{B}_{2}).
\end{eqnarray}

The end-to-end information rate from source node $j$ to source
node $i$, denoted as $R_{ji}(\mathbf{B}_{i}, \mathbf{D}_j)$, is bounded by
\begin{eqnarray}\label{e:eqbcRi}
R_{ji}(\mathbf{B}_{i}, \mathbf{D}_j) = \frac{1}{2} \min \{
\hat{R}_{{\rm{r}}i} (\mathbf{B}_{i}), \bar{R}_{j{\rm{r}}}
(\mathbf{D}_j)\}
\end{eqnarray}
where
\begin{eqnarray}
\bar{R}_{j{\rm{r}}}(\mathbf{D}_j) = \log{| \mathbf{I} +
(\mathbf{H}_{j{\rm{r}}} \mathbf{D}_j \mathbf{H}_{j{\rm{r}}}^{\rm
H}) (\sigma_{{\rm{r}}}^2)^{-1}|}. \label{e:Rbar}
\end{eqnarray}

Then the sum-rate for communication over both MA and BC phases for the
considered DF TWR can be written as \cite{TWR_PRT}
\begin{equation}\label{e:eqRtw}
R^{\rm tw}(\mathbf{B}, \mathbf{D}) = \frac{1}{2}\min\{R^{\rm
ma}(\mathbf{D}), R(\mathbf{B}, \mathbf{D})\}
\end{equation}
where \begin{eqnarray}\label{e:eqRbd}
R (\mathbf{B},\mathbf{D})= \min\{\hat{R}_{{\rm{r}}1}
(\mathbf{B}_{1}), \bar{R}_{2{\rm{r}}} (\mathbf{D}_2)\}  \qquad \nonumber\\
\qquad +\min\{\hat{R}_{{\rm{r}}2} (\mathbf{B}_{2}\!), \bar{R}_{1{\rm{r}}}
(\mathbf{D}_1\!)\}.
\end{eqnarray}

Denote the singular value decomposition (SVD) of $\mathbf{H}_{{\rm r}i}$ as $\mathbf{H}_{{\rm r}i}=\mathbf{U}_{{\rm r}i}\mathbf{\Omega}_{{\rm r}i}\mathbf{V}_{{\rm r}i}^\mathrm{H}$. We assume that the first $\mathrm{r}_{{\rm r}i}$ $(\mathrm{r}_{{\rm r}i}\leq\min(n_i, n_{\rm r}))$ diagonal elements of $\mathbf{\Omega}_{{\rm r}i}$, denoted as $\omega_{{\rm r}i}(1),\dots, \omega_{\mathrm{r}i}({\rm r}_{{\rm r}i})$, are non-zero.
\textcolor{black}{
Since the source nodes can subtract their self-interference in the BC phase and the relay has channel knowledge of $\mathbf{H}_{{\rm{r}} i}, \forall i$, the power allocation of the relay for relaying the signal in either direction should be based on waterfilling regardless of how the relay distributes its power between relaying the signals in the two directions.}
The actual water-levels used by the relay for relaying the signal from source node $j$ to source node $i$ is denoted as $1/\lambda_i, \forall i$.  With water-level $1/\lambda_i$, $\mathbf{B}_i$ can be given as $\mathbf{B}_i=\mathbf{V}_{\mathrm{r}i}\mathbf{P}_{\mathrm{r}i} (\lambda_i)\mathbf{V}_{\mathrm{r}i}^\mathrm{H}$ where $\mathbf{P}_{\mathrm{r}i}(\lambda_i)=\text{diag}\bigg(\big(\frac{1}{\lambda_i} - \frac{1}{\alpha_i(1)}\big)^{+}, \dots, \big(\frac{1}{\lambda_i}\!-\!\frac{1}{\alpha_i(\mathrm{r}_{\mathrm{r}i})}\big)^{+}, 0,\dots, 0\bigg)$

in which $\text{diag}(\cdot)$ stands for making a diagonal matrix using the given elements, $(\cdot)^{+}$ stands for projection to the positive orthant, $\alpha_i(k) = {|\omega_{{\rm r}i}(k)|^2}/{\sigma_i^2}$, and there are $(n_\mathrm{r}\!-\mathrm{r}_{\mathrm{r}i})$ zeros on the main diagonal of $\mathbf{P}_{\mathrm{r}i}(\lambda_i)$.\textcolor{black}{\footnote{\textcolor{black}{Details on waterfilling based solution of power allocation can be found, for example, in Section III.A in \cite{WF_MAC}.}}}
\textcolor{black}{It holds that
\begin{subequations}%\label{e:waterlbd}
\begin{align}
&\hat{R}_{r{\rm 1}}(\mathbf{B}_1)=\sum\limits_{k\in\mathcal{I}_1} \log\bigg(1+\bigg(
\frac{1}{\lambda_1}\alpha_1(k)-1\bigg)^{+}\bigg) \label{e:waterlbd1}\\
&\hat{R}_{r{\rm 2}}(\mathbf{B}_2)=\sum\limits_{k\in\mathcal{I}_2} \log\bigg(1 +
\bigg(\frac{1}{\lambda_2}\alpha_2(k)-1\bigg)^{+}\bigg) \label{e:waterlbd2}
%&\hat{R}_{r{\rm 1}}(\mathbf{B}_1\!)\!+\!\!\hat{R}_{r{\rm 2}}(\mathbf{B}_2\!)\!=\!\sum\limits_{i}\!\sum\limits_{k\in\mathcal{I}_i}
%\!\log\!\bigg(\!1+\!\!\bigg(\!\frac{1}{\lambda_i}
%\alpha_i(k)\!-\!1\!\!\bigg)^{+}\!\bigg).\label{e:waterlbd}
\end{align}
\end{subequations}
where $\mathcal{I}_i = \{1, \dots, {\rm r}_{{\rm r}i}\}$. Therefore, the rate
$\hat{R}_{\mathrm{r}i}(\mathbf{B}_{i})$ obtained using water-level $1/\lambda_i$ is alternatively denoted as $\hat{R}_{\mathrm{r}i}(\lambda_{i})$.}

\textcolor{black}{
From equation \eqref{e:eqRtw}, it can be seen that the maximization of the sum-rate using minimum power potentially involves balancing between $R^{\rm ma}(\mathbf{D})$ and $R(\mathbf{B}, \mathbf{D})$ and between $\hat{R}_{{\rm{r}}i}(\mathbf{B}_{i})$ and $\bar{R}_{j{\rm{r}}}(\mathbf{D}_j), \forall i$. However, it is not explicit how such rate balancing affects the power allocation of the relay and the source nodes. In order to adjust the above rates through power allocation, we introduce the relative water-levels. Same as in Part~I, $1/\mu_1(\mathbf{D}_1)$, $1/\mu_2(\mathbf{D}_2)$, and $1/\mu_{\rm
ma}(\mathbf{D})$ are defined as
}
\begin{subequations}
\begin{align}
\hspace{-2mm}&\sum\limits_{k\in\mathcal{I}_2} \log\bigg(1+\bigg(
\frac{1}{\mu_1(\mathbf{D}_1)}\alpha_2(k)-1\bigg)^{+}\bigg) =
\bar{R}_{1{\rm r}}(\mathbf{D}_1)\label{e:watermu1}\\
\hspace{-2mm}&\sum\limits_{k\in\mathcal{I}_1} \log\bigg(1 +
\bigg(\frac{1}{\mu_2(\mathbf{D}_2)}\alpha_1(k)-1\bigg)^{+}
\bigg)=\bar{R}_{2{\rm r}}(\mathbf{D}_2)\label{e:watermu2} \\
\hspace{-2mm}&\sum\limits_{i}\!\sum\limits_{k\in\mathcal{I}_i}
\!\log\bigg(\!1+\bigg(\frac{1}{\mu_{\rm ma}(\mathbf{D})}
\alpha_i(k)-1\bigg)^{+}\!\bigg)\!=R^{\rm
ma}(\mathbf{D}).\label{e:watermum}
\end{align}
\end{subequations}
\textcolor{black}{
Given the above definition, if waterfilling is performed on $\omega_{\mathrm{r}j}(k)$'s, $\forall
k\in \mathcal{I}_j$ using the water-level $1/\mu_i(\mathbf{D}_i)$,
then the information rate of the transmission from the relay to
source node $j$ using the resulting power
allocation achieves precisely $\bar{R}_{i\mathrm{r}}
(\mathbf{D}_i)$. If waterfilling is performed on
$\omega_{\mathrm{r}i}(k)$'s, $\forall k\in \mathcal{I}_i, \forall i$
using the water-level $1/\mu_{\rm ma}(\mathbf{D})$, then the
sum-rate of the transmission from the relay to the two source
nodes using the resulting power allocation
achieves precisely ${R}^{\rm ma}(\mathbf{D})$.
}
For brevity, $\mu_1(\mathbf{D}_1)$, $\mu_2(\mathbf{D}_2)$, and $\mu_{\rm ma}(\mathbf{D})$ are denoted hereafter as $\mu_1$, $\mu_2$ and $\mu_\mathrm{ma}$, respectively. The same markers/superscripts on $\mathbf{D}_i$ and/or $\mathbf{D}$ are used on $\mu_i$ and/or $\mu_\mathrm{ma}$ to represent the connection. \textcolor{black}{For example, $\mu_i(\mathbf{D}_i^{0})$ and $\mu_\mathrm{ma}(\tilde{\mathbf{D}})$ are briefly denoted as
$\mu_i^{0}$ and $\tilde{\mu}_\mathrm{ma}$, respectively.}

For the network optimization scenario considered here, the relay and the source nodes jointly maximize the sum-rate in \eqref{e:eqRtw} with minimum total transmission power in the network.\footnote{The term `sum-rate' by default means $R^{\rm tw}(\mathbf{B}, \mathbf{D})$ when we do not specify it to be the sum-rate of the BC or MA phase.} Similar to the relay optimization scenario, the relay needs to know $\mathbf{W}_1$ and $\mathbf{W}_2$ while both source nodes need to know $\mathbf{T}_{\mathrm{r}1}$ and $\mathbf{T}_{\mathrm{r}2}$. It is preferable that the TWR is able to operate in a centralized mode in which the relay can serve as a central node that carries out the computations. If the system works in a decentralized mode, it may lead to high overhead because of the information exchange during the iterative optimization process.

Given the above system model, we next solve the network optimization problem.

\section{Network optimization}\label{s:netopt}
In the network optimization scenario, the relay and the source nodes jointly optimize their power allocation to achieve sum-rate maximization with minimum total power consumption in the system for the MIMO DF TWR. Compared to the optimal solution of the relay optimization problem in Part~I, the optimal solution of the network optimization problem achieves larger sum-rate and/or less power consumption at the cost of higher computational complexity.

The sum-rate maximization part can be formulated as the following
optimization problem\footnote{The positive semi-definite
constraints $\mathbf{D}_i\succeq0, \forall i$ and
$\mathbf{B}_i\succeq0, \forall i$ are assumed as default and
omitted for brevity in all formations of optimization problems in
this paper.}
\begin{subequations}\label{e:SumR}
\begin{align}
&\mathop{\mathbf{max}}\limits_{\{\mathbf{B}, \mathbf{D}\}}  \quad R^{\rm tw}(\mathbf{B}, \mathbf{D}) \\
&\quad\mathbf{s.t.}  \quad\; \text{Tr}\{\mathbf{D}_i\}\leq P_i^{\mathrm{max}}, \forall i \\
&\qquad\qquad\!\text{Tr}\{\mathbf{B}_{1}+\mathbf{B}_{2}\}\leq P_\mathrm{r}^{\mathrm{max}}
\end{align}
\end{subequations}
where $P_i^{\mathrm{max}}$ and $P_\mathrm{r}^{\mathrm{max}}$ are the power limits for source node $i$ and the relay, respectively. The above problem is a convex problem which can be rewritten into the standard form by introducing variables $t, t_1, t_2$ as
follows
\begin{subequations}\label{e:STAN}
\begin{align}
&\hspace{-4mm}\mathop{\mathbf{max}}\limits_{\{t, t_1, t_2, \mathbf{B},
\mathbf{D}\}} \qquad t \\
&\hspace{1mm}\;\mathbf{s.t.} \quad t\leq R^{\rm ma}(\mathbf{D}), \; t\leq t_1+t_2 \\
&\hspace{1mm}\qquad\quad\! t_i \leq  \hat{R}_{\mathrm{r}j}(\mathbf{B}_j),\; t_i \leq  \bar{R}_{i\mathrm{r}}(\mathbf{D}_i),\forall i \\
&\hspace{1mm}\qquad\quad\! \text{Tr}\{\mathbf{D}_i\} \leq P_i^\mathrm{max}, \forall i, \; \text{Tr}\{\mathbf{B}_{1} + \mathbf{B}_{2}\} \leq  P_\mathrm{r}^\mathrm{max}.
\end{align}
\end{subequations}

\textcolor{black}{
If transmission power minimization is also taken into account, the following constraints are necessary
}
\begin{subequations}
\begin{align}
&\hat{R}_{\mathrm{r}i}(\mathbf{B}_{i})\leq \bar{R}_{j\mathrm{r}}
(\mathbf{D}_j), \forall i. \label{e:excons2}\\
&\;\,R^{\rm ma}(\mathbf{D})=R(\mathbf{B},\mathbf{D}). \label{e:excons1}
\end{align}
\end{subequations}
\textcolor{black}{
The reason why the above constraints are necessary if transmission power minimization
also needs to be taken into account is as follows. Given the fact that $R^{\rm ma}(\mathbf{D})<\bar{R}_{1\mathrm{r}}(\mathbf{D}_1)+\bar{R}_{2\mathrm{r}}(\mathbf{D}_2)$ whenever $\text{Tr}\{\mathbf{D}_1\}+\text{Tr}\{\mathbf{D}_2\}>0$, it can be shown that the power consumption of the relay can be reduced by reducing $\text{Tr}\{\mathbf{B}_i\}$ without decreasing the sum-rate $R^{\rm tw}(\mathbf{B}, \mathbf{D})$ in \eqref{e:eqRtw} if $\hat{R}_{\mathrm{r}i}(\mathbf{B}_{i})> \bar{R}_{j\mathrm{r}}
(\mathbf{D}_j)$. Therefore, the constraint \eqref{e:excons2} is necessary. Subject to \eqref{e:excons2}, $R^{\rm tw}(\mathbf{B}, \mathbf{D})$ in \eqref{e:eqRtw} can be written as $\min\{R^{\rm ma}(\mathbf{D}), \hat{R}_{\mathrm{r}1}(\mathbf{B}_{1})+\hat{R}_{\mathrm{r}2}(\mathbf{B}_{2})\}/2$. Using the fact that $R^{\rm ma}(\mathbf{D})< \hat{R}_{\mathrm{r}1}(\mathbf{B}_{1})+\hat{R}_{\mathrm{r}2}(\mathbf{B}_{2})$ when $\hat{R}_{\mathrm{r}1}(\mathbf{B}_{1})=\bar{R}_{2\mathrm{r}}(\mathbf{D}_2)$ and $\hat{R}_{\mathrm{r}2}(\mathbf{B}_{2})= \bar{R}_{1\mathrm{r}}(\mathbf{D}_1)$, it can be shown that the power consumption of at least one source node can be reduced without decreasing $R^{\rm tw}(\mathbf{B}, \mathbf{D})$ if $R^{\rm ma}(\mathbf{D})> R(\mathbf{B},\mathbf{D})$ while the power consumption of the relay can be reduced without decreasing $R^{\rm tw}(\mathbf{B}, \mathbf{D})$ if $R^{\rm ma}(\mathbf{D})< R(\mathbf{B}, \mathbf{D})$. Thus, the constraint \eqref{e:excons1} is also necessary.
}

Considering the constraints \eqref{e:excons2} and \eqref{e:excons1}, the problem of finding the
optimal power allocation already becomes nonconvex. \textcolor{black}{Relating \eqref{e:watermu1}-\eqref{e:watermum} with \eqref{e:waterlbd1}-\eqref{e:waterlbd2}}, the above two constraints \eqref{e:excons2} and \eqref{e:excons1} can be rewritten as
\begin{subequations}\label{e:WLconsNT}
\begin{align}
\lambda_i \geq \mu_j, \forall i \quad\quad\quad
\quad\quad\quad\;\,
\label{e:WLcons1NT} \\
\sum\limits_{i}\sum\limits_{k\in\mathcal{I}_{i}} \!\log\! \bigg(
\!1\!+\!\bigg(\frac{1}{\lambda_i}\alpha_i(k)-\!1 \bigg)^{+}
\!\bigg) \quad\quad\quad\quad\quad\quad\quad\;\, \nonumber\\
= \sum\limits_{i} \sum\limits_{k\in\mathcal{I}_{i}}
\!\log\!\bigg(\!1\! +\!\bigg( \frac{1}{\mu_\mathrm{ma}}\alpha_i(k) - \!1\bigg)^{+} \!\bigg).\label{e:WLcons2NT}
\end{align}
\end{subequations}

\textcolor{black}{It should be noted that the constraints \eqref{e:excons2} and \eqref{e:excons1}, or equivalently \eqref{e:WLcons1NT} and \eqref{e:WLcons2NT}, are not sufficient in general. Due to the intrinsic complexity of the considered problem, it is too complicated to formulate the general sufficient and necessary condition for optimality for the original problem of sum-rate maximization with minimum power consumption. Instead, we will show the sufficient and necessary optimality condition for the equivalent problems in the subcases in which the original problem can be transferred into equivalent convex problems. For other subcases, we will develop important properties based on the above necessary conditions which can significantly reduce the computational complexity of searching for the optimal solution.
}

In the scenario of network optimization, the three nodes aim at
finding the optimal matrices $\mathbf{D}$ and $\mathbf{B}$ that
minimize
%\begin{equation}
$\text{Tr} \{ \mathbf{D}_1\}+\text{Tr}\{\mathbf{D}_2\} + \text{Tr}
\{ \mathbf{B}_{1} + \mathbf{B}_{2}\}$
%\end{equation}
among all $\mathbf{D}$ and $\mathbf{B}$ that achieve the
maximum of the objective function in \eqref{e:SumR}. Considering
the fact that the optimal $\mathbf{B}$ and $\mathbf{D}$ depend on
each other, solving the considered problem generally
involves alternative optimization of $\mathbf{B}$ and $\mathbf{D}$. It is of
interest to avoid such alternative process, when it is possible, due
to its high complexity. Next we use an initial power allocation\textcolor{black}{\footnote{\textcolor{black}{
Note that the initial power allocation is not the solution to the considered problem and it is only used for enabling classification.}}} to
classify the problem of finding the optimal $\mathbf{B}$ and
$\mathbf{D}$ for network optimization into two cases, each with
several subcases. 

Consider the following initial power allocation of the source
nodes and the relay, which decides the maximum achievable sum-rates of the MA and BC phases, respectively. The source nodes solve the following problem
\begin{subequations}\label{e:MacR}
\begin{align}
&\mathop{\mathbf{max}}\limits_{\mathbf{D}} \quad R^{\rm ma}
(\mathbf{D})  \\
&\;\;\mathbf{s.t.}  \quad\;
\text{Tr}\{\mathbf{D}_i\}\leq P_i^{\rm max}, \forall i.\label{e:MacRcons}
\end{align}
\end{subequations}
It is worth mentioning that the problem \eqref{e:MacR} is a basic
power allocation problem on multiple-access channels studied in
\cite{WF_MAC}. Denote the optimal solution of the above problem as $\mathbf{D}^0=[\mathbf{D}_1^0,
\mathbf{D}_2^0]$. The relay allocates $P_\mathrm{r}^\mathrm{max}$
on $\alpha_i(k)$'s, $\forall k\in \mathcal{I}_i, \forall i$ based on
the waterfilling procedure. Denote the initial water level as
${1}/{\lambda^{0}}$. The case when $R^{\rm ma}(\mathbf{D}^0)\geq
\hat{R}_{\mathrm{r}1}(\lambda^0)+\hat{R}_{\mathrm{r}2}(\lambda^0)$, \textcolor{black}{i.e., when the maximum achievable sum-rate of the MA phase is lager than or equal to that of the BC phase},
is denoted as Case I and the case when $R^{\rm ma}(\mathbf{D}^0)<
\hat{R}_{\mathrm{r}1}(\lambda^0)+\hat{R}_{\mathrm{r}2}(\lambda^0)$, \textcolor{black}{i.e., when the maximum achievable sum-rate of the MA phase is less than that of the BC phase},
is denoted as Case II. The joint optimization over $\mathbf{B}$ and
$\mathbf{D}$ will be studied in each of these cases.
The following lemma that applies to both cases is introduced for
subsequent analysis.

\textcolor{black}{\emph{Lemma~1}: Given $\mathbf{D}_1$ and $\mathbf{D}_2$ with $P_1^{\rm max}\geq\text{Tr}\{\mathbf{D}_1\}>0$ and $P_2^{\rm max}\geq \text{Tr}\{\mathbf{D}_2\} > 0$, if $1/\mu_i > 1/\mu_{\rm ma} > 1/\mu_j$, then the following two results hold true: 1). $1/\mu_{\rm ma}(\widetilde{\mathbf{D}})\leq 1/\mu_j$ where $\widetilde{\mathbf{D}}=[\widetilde{\mathbf{D}}_1, \widetilde{\mathbf{D}}_2]$ with $\widetilde{\mathbf{D}}_i=\mathbf{0}$ and $\widetilde{\mathbf{D}}_j=\mathbf{D}_j$, 2). there exists $t\in [0, 1)$ such that with $\widehat{\mathbf{D}}_i=t\mathbf{D}_i$ and $\widehat{\mathbf{D}}_j=\mathbf{D}_j$, we have $1/\mu_i(\widehat{\mathbf{D}}_i)> 1/\mu_{\rm ma}(\widehat{\mathbf{D}}) = 1/\mu_j$ where $\widehat{\mathbf{D}}=[\widehat{\mathbf{D}}_1, \widehat{\mathbf{D}}_2]$. }

\textbf{Proof}: See Subsection \ref{s:PLm1II} in Appendix.\hfill$\blacksquare$

\textcolor{black}{ Lemma~1 relates the source nodes transmit
strategy $\mathbf{D}$ with the relative water-levels $1/\mu_1,
1/\mu_2$, and $1/\mu_{\rm ma}$. It shows a range in which the
relative water-level $1/\mu_{\rm ma}$ can change when fixing
$\mathbf{D}_j$ and changing $\mathbf{D}_i$ given that $1/\mu_i >
1/\mu_{\rm ma} > 1/\mu_j$. }

\emph{Lemma~2}: The optimal solution of the network optimization problem has the following property
\begin{equation}
%\begin{subequations}
%\begin{align}
\lambda_j=\mu_i> \mu_\mathrm{ma} \quad \text{if} \quad  \lambda_i< \lambda_j \;\,\text{or}\;\, \mu_i> \mu_\mathrm{ma}. \label{e:propty1}
%\lambda_j=\mu_i> \mu_\mathrm{ma} \quad \text{if} \quad  \mu_i> \mu_\mathrm{ma}  \label{e:propty2}
%&\min\bigg\{\!\frac{1}{\lambda_1}, \frac{1}{\lambda_2}\!\bigg\}\!=\! \min \bigg\{\!\frac{1}{\mu_1}, \frac{1}{\mu_2}\!\bigg\}\!<\! \frac{1}{\mu_{\rm ma}} \;\; \text{if} \;\; \lambda_1\neq \lambda_2 \; \label{e:propty1}\\
%&\min\bigg\{\!\frac{1}{\lambda_1}, \frac{1}{\lambda_2}\!\bigg\}\!=\!\min \bigg\{\!\frac{1}{\mu_1}, \frac{1}{\mu_2}\!\bigg\} \;\, \text{if} \;\, \min\bigg\{\!\frac{1}{\mu_1}, \frac{1}{\mu_2}\!\bigg\}\! <\! \frac{1}{\mu_{\rm ma}} \; \label{e:propty2}
%%&\frac{1}{\lambda_1}=\frac{1}{\lambda_2}= \min \bigg\{\frac{1}{\mu_{\rm ma}^0}, \frac{1}{\lambda^0}\bigg\} \; \text{if} \; \min\bigg\{\frac{1}{\mu_1}, \frac{1}{\mu_2}\bigg\} \geq \frac{1}{\mu_{\rm ma}}. \qquad\label{e:propty3}
\end{equation}

\textbf{Proof}: See Subsection \ref{s:PLm2II} in Appendix.
\hfill$\blacksquare$

\textcolor{black}{
Lemma~2 develops a property of the optimal solution that follows from the constraints \eqref{e:WLcons1NT} and \eqref{e:WLcons2NT}. This property is needed for future analysis.
}

We next study the problem of maximizing $R^{\rm tw} (\mathbf{B}, \mathbf{D})$
with minimum power consumption and find the optimal power allocation for Cases I~and~II, respectively, in the following subsections.

\subsection{Finding the optimal solution in Case I, i.e.,
$R^{\rm ma}(\mathbf{D}^0) \geq \hat{R}_{\mathrm{r}1}
(\lambda^0)+\hat{R}_{\mathrm{r}2}(\lambda^0)$}

\textcolor{black}{Since $R^{\rm ma}(\mathbf{D}^0) \geq \hat{R}_{\mathrm{r}1}(\lambda^0)+\hat{R}_{\mathrm{r}2}(\lambda^0)$, it can be shown that $1/\lambda_0\leq 1/\mu_\mathrm{ma}^0$. In this case, the sum-rate $R^{\rm tw}(\mathbf{B}, \mathbf{D})$ in \eqref{e:eqRtw} is upper-bounded by the sum-rate $\hat{R}_{\mathrm{r}1} (\lambda^0)+\hat{R}_{\mathrm{r}2}(\lambda^0)$. The following two subcases should be considered separately.}

\textcolor{black}{Subcase I-1: The following convex optimization problem is feasible}
\begin{subequations}\label{e:Sub2}
\begin{align}
&\mathop{\mathbf{min}}\limits_{\mathbf{D}}\quad \text{Tr} \{ \mathbf{D}_1\}+\text{Tr}\{\mathbf{D}_2\}\label{e:Sub2Obj}\\
&\;\;\mathbf{ s.t.}  \quad\; R^{\rm ma}(\mathbf{D})\geq \hat{R}_{\mathrm{r}1}(\lambda^0)+\hat{R}_{\mathrm{r}2}(\lambda^0)\label{e:Sub2cons1}\\
&\hspace{1mm}\qquad\quad \bar{R}_{1\mathrm{r}}(\mathbf{D}_1)\geq \hat{R}_{\mathrm{r}2}(\lambda^0)\label{e:Sub2cons2}\\
&\qquad\quad\,\, \bar{R}_{2\mathrm{r}}(\mathbf{D}_2)\geq \hat{R}_{\mathrm{r}1}(\lambda^0)\label{e:Sub2cons3}\\
&\qquad\quad\,\, \text{Tr}(\mathbf{D}_i) \leq   P_i^\mathrm{max},\forall i. \label{e:Sub2cons4}
\end{align}
\end{subequations}
\textcolor{black}{In this subcase, the maximum sum-rate $R^{\rm tw}(\mathbf{B}, \mathbf{D})$ can achieve
$\hat{R}_{\mathrm{r}1} (\lambda^0)+\hat{R}_{\mathrm{r}2}
(\lambda^0)$. In order to achieve this maximum sum-rate, it is necessary that $\lambda_1=\lambda_2=\lambda^0$. Therefore, the relay should use up all available power
$P_\mathrm{r}^\mathrm{max}$ at optimality, and the optimal $\mathbf{B}_i, \forall
i$ is equal to $\mathbf{V}_{\mathrm{r}i} \mathbf{P}_{\mathrm{r}i}
(\lambda^0) \mathbf{V}_{\mathrm{r}i}^\mathrm{H}$ where
$\mathbf{P}_{\mathrm{r}i} (\lambda_i)$ is given in Section~\ref{s:sysm}. As a result, the original problem simplifies to finding the optimal $\mathbf{D}_1$ and $\mathbf{D}_2$ such that $R^{\rm tw}(\mathbf{B}, \mathbf{D})$ achieves $\hat{R}_{\mathrm{r}1} (\lambda^0)+\hat{R}_{\mathrm{r}2}(\lambda^0)$ with minimum power consumption.}
\textcolor{black}{
Using equations \eqref{e:eqRtw} and \eqref{e:eqRbd}, it can be shown that the sufficient and necessary condition for  $\mathbf{D}$ to be optimal in this subcase is that $\mathbf{D}$ is the optimal solution to the convex optimization problem \eqref{e:Sub2}. } \textcolor{black}{
Denoting the optimal solution to the problem \eqref{e:Sub2} as $\mathbf{D}^\star=[\mathbf{D}_1^\star, \mathbf{D}_2^\star]$, the total power consumption in this subcase is $P_\mathrm{r}^\mathrm{max}+\text{Tr} \{ \mathbf{D}_1^\star\}+\text{Tr}\{\mathbf{D}_2^\star\}$.}

\textcolor{black}{It can be seen that the optimal solution of $\mathbf{B}$ and $\mathbf{D}$ described above satisfies the necessary condition \eqref{e:WLcons1NT} as the constraints \eqref{e:Sub2cons2} and \eqref{e:Sub2cons3} are considered in the problem \eqref{e:Sub2}. It can also be shown that the above optimal solution in Subcase~I-~1 satisfies the necessary condition \eqref{e:WLcons2NT}, as stated in the following theorem.}

\textcolor{black}{\textbf{Theorem~1}: The optimal solution in Subcase I-1 satisfies $\mu_{\rm ma}^{\star}=\lambda^0$ where $\mu_{\rm ma}^{\star}=\mu_{\rm ma}(\mathbf{D}^{\star})$, and thereby satisfies \eqref{e:WLcons2NT} given that $\lambda_1=\lambda_2=\lambda^0$ at optimality.}

\textbf{Proof}: See Subsection \ref{s:PTh1} in Appendix.

\textcolor{black}{Considering the constraints \eqref{e:Sub2cons1}-\eqref{e:Sub2cons4}, it can be seen that the problem \eqref{e:Sub2} is feasible if and only if the optimal solution to the following problem
\begin{subequations}\label{e:Sub3}
\begin{align}
&\mathop{\mathbf{max}}\limits_{\mathbf{D}} \quad
\bar{R}_{j\mathrm{r}}(\mathbf{D}_j) \\
&\;\; \mathbf{s.t.}  \quad\;\,
\bar{R}_{i\mathrm{r}}(\mathbf{D}_i) \geq \hat{R}_{\mathrm{r}j}(\lambda^0) \\
&\qquad\quad\;\, R^{\rm ma}(\mathbf{D}) \geq \hat{R}_{\mathrm{r}1}(\lambda^0)+\hat{R}_{\mathrm{r}2}(\lambda^0)\\
&\qquad\quad\;\, \text{Tr}(\mathbf{D}_i)\!\leq \! P_i^\mathrm{max}, \forall i
\end{align}
\end{subequations}
denoted as $\mathbf{D}^{*}$, satisfies $\bar{R}_{j\mathrm{r}} (\mathbf{D}_j^{*})\geq \hat{R}_{\mathrm{r}i}(\lambda^0)$.\footnote{\textcolor{black}{Note that if $\bar{R}_{j\mathrm{r}} (\mathbf{D}_j^{*})\geq \hat{R}_{\mathrm{r}i}(\lambda^0)$ for $i=1, j=2$ in \eqref{e:Sub3} then it also holds that $\bar{R}_{j\mathrm{r}} (\mathbf{D}_j^{*})\geq \hat{R}_{\mathrm{r}i}(\lambda^0)$ for $i=2, j=1$ and vice versa.}} However, it is possible that $\bar{R}_{j\mathrm{r}} (\mathbf{D}_j^{*})< \hat{R}_{\mathrm{r}i}(\lambda^0)$. It is also possible that the problem \eqref{e:Sub3} is not even feasible. In both of the above two situations the problem \eqref{e:Sub2} is infeasible. This leads to the second subcase of Case I.}

\textcolor{black}{Subcase I-2: The problem \eqref{e:Sub2} is infeasible.}

\textcolor{black}{
Unlike Subcase I-1, the maximum sum-rate $R^{\rm tw}(\mathbf{B}, \mathbf{D})$ in this subcase cannot achieve $\hat{R}_{\mathrm{r}1} (\lambda^0)+\hat{R}_{\mathrm{r}2}(\lambda^0)$. As mentioned above, there are two possible situations when the problem \eqref{e:Sub2} is infeasible: (i) $\bar{R}_{j\mathrm{r}} (\mathbf{D}_j^{*})< \hat{R}_{\mathrm{r}i}(\lambda^0)$, and (ii) the problem \eqref{e:Sub3} is infeasible. Using Lemma~1 in Part~I of this two-part paper \cite{PartI} and the fact that $R^{\rm ma}(\mathbf{D}^0) \geq \hat{R}_{\mathrm{r}1}
(\lambda^0)+\hat{R}_{\mathrm{r}2}(\lambda^0)$ for Case~I, it can be shown that if the problem \eqref{e:Sub3} is infeasible for specific values of $i$ and $j$, then it is feasible (but $\bar{R}_{j\mathrm{r}} (\mathbf{D}_j^{*})< \hat{R}_{\mathrm{r}i}(\lambda^0)$) when the values of $i$ and $j$ are switched. Therefore, the problem \eqref{e:Sub2} is infeasible if and only if there exists at least one specific value of $j$ in $\{1,2\}$ such that $\bar{R}_{j\mathrm{r}} (\mathbf{D}_j^{*})< \hat{R}_{\mathrm{r}i}(\lambda^0)$ in the problem \eqref{e:Sub3}. It infers, based on the definitions \eqref{e:watermu1}-\eqref{e:watermum}, that $1/\mu_j< 1/\lambda^0$ whenever $1/\mu_\mathrm{ma} \geq 1/\lambda^0$ and
$1/\mu_i \geq 1/\lambda^0$. As a result, whenever $1/\mu_\mathrm{ma} \geq 1/\lambda^0$, or equivalently, $R^{\rm ma}(\mathbf{D}) \geq \hat{R}_{\mathrm{r}1}(\lambda^0)+\hat{R}_{\mathrm{r}2}(\lambda^0)$, the sum-rate $R^{\rm tw}(\mathbf{B}, \mathbf{D})$ is bounded by $\hat{R}_{\mathrm{r}1}(\lambda_1)+\hat{R}_{\mathrm{r}2}(\lambda_2)$ (according to equation \eqref{e:eqRtw}), which is less than $\hat{R}_{\mathrm{r}1}(\lambda^0)+\hat{R}_{\mathrm{r}2}(\lambda^0)$ when $1/\mu_j< 1/\lambda^0$ (according to the constraint \eqref{e:WLcons1NT}). Moreover, whenever $1/\mu_\mathrm{ma} < 1/\lambda^0$, or equivalently, $R^{\rm ma}(\mathbf{D}) < \hat{R}_{\mathrm{r}1}(\lambda^0)+\hat{R}_{\mathrm{r}2}(\lambda^0)$, the sum-rate $R^{\rm tw}(\mathbf{B}, \mathbf{D})$ is bounded by $R^{\rm ma}(\mathbf{D})$ (according to equation \eqref{e:eqRtw}), which is also less than $\hat{R}_{\mathrm{r}1}(\lambda^0)+\hat{R}_{\mathrm{r}2}(\lambda^0)$. Therefore, the maximum sum-rate $R^{\rm tw}(\mathbf{B}, \mathbf{D})$ in this subcase cannot achieve $\hat{R}_{\mathrm{r}1} (\lambda^0)+\hat{R}_{\mathrm{r}2}(\lambda^0)$.
}

\textcolor{black}{We specify $j$ for this subcase so that the problem \eqref{e:Sub3} is feasible but $\bar{R}_{j\mathrm{r}} (\mathbf{D}_j^{*})< \hat{R}_{\mathrm{r}i}(\lambda^0)$}. The following theorem characterizes the
optimal solution in this subcase.

\textbf{Theorem~2}: Denote the optimal $\mathbf{D}_l$ in Subcase
I-2 as $\mathbf{D}_l^*, \forall l\in\{1, 2\}$ and the optimal $\lambda_l$ as
$\lambda_l^*, \forall l$. The optimal strategies for the source
nodes and the relay satisfy the following properties:
\begin{itemize}
\item [1.] $\min\limits_{l}\{1/\mu_l^*\}<1/\mu_\mathrm{ma}^* <
1/\lambda^0$;
\item [2.] The relay uses full power $P_\mathrm{r}^\mathrm{max}$;
\item [3.] $\mathbf{D}^*$ maximizes $\min\limits_l \{1/\mu_l\}$ among all $\mathbf{D}$'s that satisfy
\begin{subequations}\label{e:ProfTh3cons}
\begin{align}
&R^{\rm ma}(\mathbf{D}) \geq  R^{\rm ma}(\mathbf{D}^*)\\
&\;\;\text{Tr}(\mathbf{D}_l) \leq  P_l^\mathrm{max}, \forall l
\end{align}
\end{subequations}
\item [4.] $1/\mu_j^*<1/\mu_i^*$.
\end{itemize}

\textbf{Proof}:  Please see Subsection~\ref{s:PTh2} in Appendix.

\textcolor{black}{
While the original problem cannot be simplified into an equivalent form in this subcase, the properties in the above theorem
help to significantly reduce the complexity of searching for the optimal solution by narrowing down the set of qualifying power allocations.}
Denote the $\mathbf{D}_j$ that maximizes $\bar{R}_{j\mathrm{r}}
(\mathbf{D}_j)$ subject to the constraints $\mu_j\geq \mu_{\rm ma}$
and $\text{Tr}\{\mathbf{D}_j\}\leq P_j^\mathrm{max}$ as
$\mathbf{D}_j^{\mathrm{l}}$ and the corresponding $\mu_j$  as
$\mu_j^{\mathrm{l}}$. According to Theorem~2, if
$\hat{R}_{\mathrm{r}i}(\bar{\lambda}_i)+
\hat{R}_{\mathrm{r}j}(\bar{\lambda}_j) \leq R^{\rm
ma}(\bar{\mathbf{D}})$, where
\begin{subequations}\label{e:oneshotcons}
\begin{align}
\bar{\lambda}_i=\mu_j^{\mathrm{l}}\qquad\qquad\qquad\\
\text{Tr} \{\mathbf{P}(\bar{\lambda}_i) \} + \text{Tr} \{ \mathbf{P}
(\bar{\lambda}_j) \} = P_\mathrm{r}^\mathrm{max}
\end{align}
\end{subequations}
and $\bar{\mathbf{D}}$ is the optimal solution of the following problem
\begin{subequations}
\begin{align}
&\mathop{\mathbf{max}}\limits_{\mathbf{D}} \quad R^{\rm ma} (\mathbf{D}) \qquad\qquad\quad\\
&\;\;\, \mathbf{s.t.} \quad\, \bar{R}_{j\mathrm{r}}(\mathbf{D}_j) \geq \bar{R}_{j\mathrm{r}}(\mathbf{D}_j^\mathrm{l})\\
&\qquad\quad\;\, \text{Tr}(\mathbf{D}_l)\!\leq \! P_l^\mathrm{max}, \forall l
\end{align}
\end{subequations}
then the optimal $\mathbf{B}_l, \forall l$ in Subcase I-2 is given by
$\mathbf{B}_l=\mathbf{V}_{\mathrm{r}l}\mathbf{P}_{\mathrm{r}l}
(\bar{\lambda}_l)\mathbf{V}_{\mathrm{r}l}^\mathrm{H}$ and the
optimal $\mathbf{D}$ is the solution to the following power
minimization problem
\begin{subequations}\label{e:CI3oneshot}
\begin{align}
&\mathop{\mathbf{min}}\limits_{\mathbf{D}}\quad \text{Tr} \{ \mathbf{D}_1\}+\text{Tr}\{\mathbf{D}_2\}\\
&\;\, \mathbf{s.t.}  \quad\; R^{\rm ma}(\mathbf{D})\geq \sum\limits_l \hat{R}_{\mathrm{r}l}(\bar{\lambda}_l)\\
&\qquad\quad\, \bar{R}_{i\mathrm{r}}(\mathbf{D}_i)\geq \hat{R}_{\mathrm{r}j}(\bar{\lambda}_j)\\
&\qquad\quad\, \bar{R}_{j\mathrm{r}}(\mathbf{D}_j)\geq \hat{R}_{\mathrm{r}i}(\bar{\lambda}_i)\\
&\qquad\quad\, \text{Tr}(\mathbf{D}_l) \!\leq \! P_l^\mathrm{max}, \forall l.
\end{align}
\end{subequations}

If $\hat{R}_{\mathrm{r}i}(\bar{\lambda}_i)+ \hat{R}_{\mathrm{r}j}
(\bar{\lambda}_j) > R^{\rm ma}(\bar{\mathbf{D}})$, then according
to Theorem~2, the optimal solution can be found by maximizing the
objective $R^{\rm tw}(\mathbf{B},\mathbf{D})$, denoted as $R^{\rm
obj}$, that can be achieved by both $R^{\rm ma}(\mathbf{D})$ and
$\sum\limits_l\hat{R}_{\mathrm{r}l}(\lambda_l)$ subject to the following two
constraints: 1). $1/\lambda_{i}=1/\tilde{\mu}_j$ (according to
Lemma~2, Properties 1 and 4 of Theorem 2); 2). $1/\lambda_{j}$
is obtained by waterfilling the remaining power on
$\alpha_{\mathrm{r}j}(k), \forall k\in \mathcal{I}_j$ (Property 2
of Theorem 2), where $1/\tilde{\mu}_j$ is the optimal value of the
objective function in the following optimization problem (Property
3 of Theorem 2)
\begin{subequations}\label{e:maxmuj}
\begin{align}
&\mathop{\mathbf{max}}\limits_{\mathbf{D}}\quad \frac{1}{\mu_j}\\
&\;\,\mathop{\mathbf{s.t.}}\quad\, R^{\rm ma}(\mathbf{D})\geq R^{\rm obj} \label{e:maxmujcons1}\\
&\qquad\quad\;\,\text{Tr}(\mathbf{D}_l)\leq P_l^\mathrm{max}, \forall l.\label{e:maxmujcons2}
\end{align}
\end{subequations}
Since maximizing $1/\mu_j$ is equivalent to maximizing
$\bar{R}_{j\mathrm{r}} (\mathbf{D}_j)$, the objective function of
the above problem can be substituted by $\bar{R}_{j\mathrm{r}}
(\mathbf{D}_j)$, and $1/\tilde{\mu}_j$ can be obtained from the
optimal value of $\bar{R}_{j\mathrm{r}}(\mathbf{D}_j)$ in the
above problem using \eqref{e:watermu1} or \eqref{e:watermu2}. \textcolor{black}{As mentioned at the beginning of Subcase I-2}, the optimal $R^{\rm tw}(\mathbf{B},
\mathbf{D})$ is less than $\sum\limits_l \hat{R}_{\mathrm{r}l}
(\lambda^0)$. Therefore, starting from the point by setting
$R^{\rm obj}=\sum\limits_l \hat{R}_{\mathrm{r}l}(\lambda^0)$, we
can adjust $R^{\rm obj}$ to achieve the optimal $R^{\rm
tw}(\mathbf{B}, \mathbf{D})$ by solving the following problem
\begin{subequations}\label{e:minwl}
\begin{align}
&\mathop{\mathbf{max}}\limits_{\mathbf{D}}\quad\bar{R}_{j\mathrm{r}}(\mathbf{D}_j)\\
&\;\,\mathop{\mathbf{s.t.}}\quad\, R^{\rm ma}(\mathbf{D}) \geq  R^{\rm obj} \\
&\qquad\quad\;\,\text{Tr}(\mathbf{D}_l) \leq   P_l^\mathrm{max}, \forall l
\end{align}
\end{subequations}
and obtain the resulting $1/\tilde{\mu}_j$ from the above problem.
Setting $1/\lambda_{i}=1/\tilde{\mu}_j$ and allocating all the
remaining power on $\alpha_{\mathrm{r}j}(k)$'s, $\forall k\in
\mathcal{I}_j$, if the resulting $\sum\limits_l
\hat{R}_{\mathrm{r}l}(\lambda_l)$ is less than $R^{\rm obj}$, then
$R^{\rm obj}$ should be decreased and the above process should be
repeated. If the resulting $\sum\limits_l
\hat{R}_{\mathrm{r}l}(\lambda_l)$ is larger than $R^{\rm obj}$,
then $R^{\rm obj}$ should be increased and the above process
should be repeated. The optimal solution is found when the
resulting $\sum\limits_l \hat{R}_{\mathrm{r}l}(\lambda_l)$ is
equal to $R^{\rm ma}(\mathbf{D})$. With an
appropriate step size of increasing/decreasing $R^{\rm obj}$,
$R^{\rm obj}$ converges to the optimal $R^{\rm tw}(\mathbf{B},
\mathbf{D})$ in the above procedure.

After obtaining the optimal $R^{\rm obj}$, $1/\tilde{\mu}_j$ and
$\lambda_i$, the source nodes need to solve the problem of  power
minimization, which is
\begin{subequations}\label{e:SubIfnl}
\begin{align}
&\mathop{\mathbf{min}}\limits_{\mathbf{D}}\quad \text{Tr} \{ \mathbf{D}_1\}+\text{Tr}\{\mathbf{D}_2\} \\
&\;\;\mathbf{s.t.}  \quad\, R^{\rm ma}(\mathbf{D})\geq R^{\rm obj}\\
&\qquad\quad\; \bar{R}_{i\mathrm{r}}(\mathbf{D}_i)\geq \hat{R}_{\mathrm{r}j}(\lambda_j)\\
&\qquad\quad\; \bar{R}_{j\mathrm{r}}(\mathbf{D}_j)\geq \hat{R}_{\mathrm{r}i}(\lambda_i)\\
&\qquad\quad\; \text{Tr}(\mathbf{D}_l)\leq P_l^\mathrm{max}, \forall l.
\end{align}
\end{subequations}
However, it can be shown that if $\bar{R}_{j\mathrm{r}}
(\tilde{\mathbf{D}}_j)$ is not the maximum that $\bar{R}_{j\mathrm{r}}
(\mathbf{D}_j)$ can achieve subject to the constraint
\eqref{e:maxmujcons2} (without the constraint
\eqref{e:maxmujcons1}), then $\mathbf{B}$ and $\mathbf{D}$ remain the
same after solving the above problem.

Using Property~2 of Theorem~2, it can be seen from \eqref{e:CI3oneshot} and \eqref{e:SubIfnl} that the minimization of total power consumption becomes the minimization of the source node power consumption in Subcase I-2 since the relay always needs to consume all its available power for achieving optimality.

The complete procedure of finding the optimal solution in Case~I is summarized in the algorithm in Table~\ref{t:Al_I3}. \textcolor{black}{The algorithm finds the optimal solution either in one shot (Steps~1~and~2) or through a bisection search for the optimal $R^{\rm obj}$ (Steps 3 to 5). Denoting $\Delta=R^{\mathrm{max}}-R^{\mathrm{min}}$, the worst case number of iterations in the bisection search is $\log(\Delta/\epsilon)$. Within each iteration, a convex problem, i.e., problem \eqref{e:minwl}, is solved followed by a simple waterfilling procedure (linear complexity) for the given $R^{\rm obj}$. Therefore, the complexity of the proposed algorithm is low.}

\begin{table}
\begin{center}
\caption {Algorithm for finding the optimal solution for Case~I}\label{t:Al_I3}
\vspace{-3mm}
\begin{tabular*}{0.49\textwidth}[t]{p{0.48\textwidth}}
\hline\hline %{\sigma_{ni}^2}/
1. Check if the problem \eqref{e:Sub2} is feasible. If yes, find
the optimal $\mathbf{D}$ from the problem \eqref{e:Sub2}. The optimal
$\mathbf{B}$ is given by $\mathbf{B}_i=\mathbf{V}_{\mathrm{r}i}
\mathbf{P}_{\mathrm{r}i}(\lambda^0)\mathbf{V}_{\mathrm{r}i}^\mathrm{H},
\forall i$. Otherwise, specify $j$ so that the problem \eqref{e:Sub3} is feasible but $\bar{R}_{j\mathrm{r}} (\mathbf{D}_j^{*})< \hat{R}_{\mathrm{r}i}(\lambda^0)$ and proceed to Step~2.\\
\hline 2. Obtain $\mathbf{D}_j^{\mathrm{l}}$ and $\mu_j^{\mathrm{l}}$.
Calculate $\bar{\lambda}_l, \forall l$ using \eqref{e:oneshotcons}.
Check if $\sum\limits_l \hat{R}_{\mathrm{r}l}(\bar{\lambda}_l) \leq
R^{\rm ma}(\bar{\mathbf{D}})$. If yes, the optimal $\mathbf{B}$
is given by $\mathbf{B}_l=\mathbf{V}_{\mathrm{r}l}\mathbf{P}_{\mathrm{r}
l}(\bar{\lambda}_l)\mathbf{V}_{\mathrm{r}l}^H, \forall l$. Find the
optimal $\mathbf{D}$ from \eqref{e:CI3oneshot}. Otherwise, proceed to
Step~3.\\
\hline 3. Set $R^{\mathrm{max}}=\sum\limits_l \hat{R}_{\mathrm{r}l}
(\lambda^0)$ and $R^{\mathrm{min}}=0$. Initialize $R^{\rm obj} =
R^{\mathrm{max}}$ and proceed to Step~4.\\
\hline 4. Solve the problem \eqref{e:minwl} and obtain $\mathbf{D}$
and $1/\tilde{\mu}_j$. Set $1/\lambda_{i}=1/\tilde{\mu}_j$.  Allocate
all the remaining power on $\alpha_{\mathrm{r}j}(k)'s, \forall k \in
\mathcal{I}_j$ using waterfilling and obtain $1/\lambda_j$. Check if
$|\sum\limits_l \hat{R}_{\mathrm{r}l}(\lambda_l)-R^{\rm ma}(\mathbf{D})
| < \epsilon$, where $\epsilon$ is the positive tolerance. If yes,
proceed to Step~6 with $R^{\rm obj}$ and $\lambda_l, \forall l$. Otherwise,
proceed to Step~5.\\
\hline 5. If $R^{\rm ma}(\mathbf{D})-\sum\limits_l \hat{R}_{\mathrm{r}
l} (\lambda_l)> \epsilon$, set $R^{\rm max}=R^{\rm obj}$. If $\sum\limits_l
\hat{R}_{\mathrm{r}l}(\lambda_l)-R^{\rm ma}(\mathbf{D})> \epsilon$, set
$R^{\rm min}=R^{\rm obj}$. Let $R^{\rm obj}=(R^{\rm max}+R^{\rm min})/2$
and go back to Step~4.\\
\hline 6. Solve the power minimization problem \eqref{e:SubIfnl}. Output
$\mathbf{D}$ and $\mathbf{B}_l=\mathbf{V}_{\mathrm{r}l}
\mathbf{P}_{\mathrm{r}l}(\lambda_l)\mathbf{V}_{\mathrm{r}l}^\mathrm{H},
\forall l$.
\\
\hline \hline
\end{tabular*}
\end{center}
\vspace{-0.5cm}
\end{table}

\textcolor{black}{
Subcases I-1 and I-2 cover all possible situations for Case I that $R^{\rm ma}(\mathbf{D}^0) \geq \hat{R}_{\mathrm{r}1}(\lambda^0)+ \hat{R}_{\mathrm{r}2}(\lambda^0)$.
}

\subsection{Finding the optimal solution in Case II, i.e., $R^{\rm ma}
(\mathbf{D}^0) < \hat{R}_{\mathrm{r}1}(\lambda^0)+\hat{R}_{\mathrm{r}2}
(\lambda^0)$}

Since $R^{\rm ma}(\mathbf{D}^0) < \hat{R}_{\mathrm{r}1}
(\lambda^0)+\hat{R}_{\mathrm{r}2}(\lambda^0)$, it can be seen \textcolor{black}{using \eqref{e:waterlbd1}, \eqref{e:waterlbd2} and \eqref{e:watermum}} that
$1/\lambda_0>1/\mu_\mathrm{ma}^0$. The following four subcases are
possible.

Subcase II-1: $1/\mu_\mathrm{ma}^0\leq \min\{1/\mu_1^0, 1 /
\mu_2^0\}$. In this subcase, the maximum $R^{\rm tw}(\mathbf{B},
\mathbf{D})$ is bounded by $R^{\rm ma}(\mathbf{D}^0)$. The optimal $\mathbf{D}$  is
$\mathbf{D}^0$, and consequently both source nodes use all their available power at optimality.
\textcolor{black}{
It can be shown that the sufficient and necessary condition for $\mathbf{B}$ to be optimal in this subcase is that $\mathbf{B}$ is the optimal solution to the following convex optimization problem
}
\begin{subequations}\label{e:SubII1}
\begin{align}
&\mathop{\mathbf{min}}\limits_{\mathbf{B}} \quad \text{Tr} \{
\mathbf{B}_{1}+\mathbf{B}_{2}\}\\
&\;\; \mathbf{s.t.} \;\;\;\,\, \hat{R}_{\mathrm{r}1}(\mathbf{B}_{1})+\hat{R}_{\mathrm{r}2}(\mathbf{B}_{2})\geq R^{\rm ma}(\mathbf{D}^0).
\end{align}
\end{subequations}
The solution of \eqref{e:SubII1} can be found in closed-form and
it is given by $\mathbf{B}_i = \mathbf{V}_{\mathrm{r}i}
\mathbf{P}_{\mathrm{r}i} (\mu_\mathrm{ma}^0)
\mathbf{V}_{\mathrm{r}i}^\mathrm{H}, \forall i$.

Subcase II-2: \textcolor{black}{there exist $i$ and $j$ such that} $1/\mu_j^0 \leq 1/\mu_\mathrm{ma}^0 < 1 / \mu_i^0
\leq1/\lambda^0$. In this subcase, the maximum $R^{\rm tw}(\mathbf{B},
\mathbf{D})$ is also bounded by $R^{\rm ma}(\mathbf{D}^0)$. Therefore, the optimal $\mathbf{D}$ is
$\mathbf{D}^0$ and both source nodes use all their available power at optimality.
\textcolor{black}{
It can be shown that the sufficient and necessary condition for $\mathbf{B}$ to be optimal in this subcase is that $\mathbf{B}$ is the optimal solution to the following convex optimization problem
}
\begin{subequations}\label{e:SubII2}
\begin{align}
&\mathop{\mathbf{min}}\limits_{\mathbf{B}} \quad \text{Tr} \{
\mathbf{B}_{1} +\mathbf{B}_{2}\}\\
&\;\; \mathbf{s.t.} \quad\; \hat{R}_{\mathrm{r}1}(\mathbf{B}_{1})+\hat{R}_{\mathrm{r}2}
(\mathbf{B}_{2}) \geq R^{\rm ma}(\mathbf{D}^0) \\
&\qquad\quad\,\, \hat{R}_{\mathrm{r}i} (\mathbf{B}_{i})= \bar{R}_{j\mathrm{r}}
(\mathbf{D}_j^0).
\end{align}
\end{subequations}
The solution of \eqref{e:SubII2} can also be expressed in
closed-form. The optimal $\mathbf{B}_{i}$ is given by
$\mathbf{B}_i = \mathbf{V}_{\mathrm{r}i} \mathbf{P}_{\mathrm{r}i}
(\mu_j^0)\mathbf{V}_{\mathrm{r}i}^\mathrm{H}$ and the optimal
$\mathbf{B}_{j}$ is given by $\mathbf{B}_j =
\mathbf{V}_{\mathrm{r}j} \mathbf{P}_{\mathrm{r}j}
(\lambda_j)\mathbf{V}_{\mathrm{r}j}^H$, where $\lambda_j$
satisfies $\hat{R}_{\mathrm{r}j}(\lambda_j)=R^{\rm
ma}(\mathbf{D}^0)-\bar{R}_{j\mathrm{r}}(\mathbf{D}_j^0)$.

Subcase II-3: \textcolor{black}{there exist $i$ and $j$ such that} $1/\mu_j^0 \leq 1/\mu_\mathrm{ma}^0 < 1/\lambda^0 <
1/\mu_i^0$ and there exists $\lambda_j$ such that
\begin{subequations}\label{e:SubII3con}
\begin{align}
&\quad\hat{R}_{\mathrm{r}j}(\lambda_j)\geq R^{\rm ma} (\mathbf{D}^0) -
\bar{R}_{j\mathrm{r}} (\mathbf{D}_j^0)\\
&\text{Tr} \{\mathbf{P}_{\mathrm{r}j}(\lambda_j)\}\leq
P_\mathrm{r}^\mathrm{max} - \text{Tr} \{ \mathbf{P}_{\mathrm{r}i}
(\mu_j^0)\}.
\end{align}
\end{subequations}
The optimal solutions of $\mathbf{B}$ and $\mathbf{D}$ in this
subcase are the same as those given in Subcase II-2.

In the above three subcases, the maximum achievable $R^{\rm tw}(\mathbf{B},
\mathbf{D})$ is $R^{\rm ma}(\mathbf{D}^0)$. Therefore, the original problem of maximizing $R^{\rm tw}(\mathbf{B},
\mathbf{D})$ with minimum total power consumption in the network simplifies to the problem that the relay uses minimum power consumption to achieve the BC phase sum-rate $\hat{R}_{\mathrm{r}1}(\mathbf{B}_{1})+\hat{R}_{\mathrm{r}2}(\mathbf{B}_{2})$ that is equal to $R^{\rm ma}(\mathbf{D}^0)$.

Subcase II-4: \textcolor{black}{there exist $i$ and $j$ such that} $1/\mu_j^0 \leq 1/\mu_\mathrm{ma}^0 < 1/\lambda^0 <
1/\mu_i^0$ and there is no $\lambda_j$ that satisfies the
conditions in \eqref{e:SubII3con}. In this subcase, the maximum
$R(\mathbf{B}, \mathbf{D})$ cannot achieve $R^{\rm
ma}(\mathbf{D}^0)$ although $R^{\rm ma}(\mathbf{D}^0)<
\hat{R}_{\mathrm{r}1}(\lambda^0)+\hat{R}_{\mathrm{r}2}(\lambda^0)$.

\textbf{Theorem~3}: Denote the optimal $\mathbf{D}_l$ as
$\mathbf{D}_l^*, \forall l$ and the optimal $\lambda_l$ as
$\lambda_l^*, \forall l$. In Subcase II-4, the optimal strategies
for the source nodes and the relay satisfy the following
properties:
\begin{itemize}
\item [1.] $\min\limits_{l}\{1/\mu_l^*\}<1/\mu_\mathrm{ma}^* <
1/\mu_\mathrm{ma}^0$;
\item [2.] Properties 2-4 in Theorem~2 also apply for Subcase II-4.
\end{itemize}

\textbf{Proof}: See Subsection~\ref{s:PTh3} in Appendix.

According to Theorem~3, the original problem of maximizing $R^{\rm tw}(\mathbf{B}, \mathbf{D})$ with minimum total power consumption becomes the problem that the source nodes and the relay jointly find the maximum achievable $R^{\rm tw}(\mathbf{B}, \mathbf{D})$ with the relay using all its available power and the source nodes using minimum power. From Theorem 3, it can be seen that the optimal solutions in the
Subcases I-2 and II-4 share very similar properties. There is also
an intuitive way to understand the similarity. Although Subcases
I-2 and II-4 are classified to opposite cases according to the
initial power allocation, it is the same for them that
$R(\mathbf{B}, \mathbf{D})$ cannot achieve $R^{\rm ma}
(\mathbf{D}^0)$. As a result, the relay needs to use as much power
as possible and the source nodes need to decrease $R^{\rm ma}
(\mathbf{D})$ from $R^{\rm ma}(\mathbf{D}^0)$ until the maximum
$R(\mathbf{B}, \mathbf{D})$ can achieve $R^{\rm ma} (\mathbf{D})$.
This similarity leads to the common properties of the above two
subcases. Moreover, due to this similarity between
Theorems~2~and~3, Steps~2~to~6 of the algorithm in Table~\ref{t:Al_I3} can be used
to derive the optimal solution in Subcase II-4 if the part of
$R^{\rm max}=\sum\limits_l\hat{R}_{\mathrm{r}l}(\lambda^0)$ in
Step~3 is substituted by $R^{\rm max} = R^{\rm ma}(\mathbf{D}^0)$.

\begin{table}[t]
\begin{center}
\caption {Summary of the overall algorithm for network optimization.}
\label{t:SumAl}
\vspace{-3mm}
\begin{tabular*}{0.49\textwidth}[t]{p{0.48\textwidth}}
\hline\hline %{\sigma_{ni}^2}/
1. \textbf{Initial power allocation}. The source nodes solve the
MA sum-rate maximization problem \eqref{e:MacR} and obtain
$\mathbf{D}^0$, $\bar{R}_{i\mathrm{r}}(\mathbf{D}_i^0), \forall i$,
and $R^{\rm ma}(\mathbf{D}^0)$. The relay obtains $\lambda^0$ and
$\hat{R}_{\mathrm{r}i}(\lambda^0), \forall i$.\\
\hline 2. \textbf{Determining the cases}. Check if $R^{\rm ma}
(\mathbf{D}^0)\geq \sum\limits_i \hat{R}_{\mathrm{r}i}(\lambda^i)$.
If yes, proceed to Step~3. Otherwise, proceed to Step~4. \\
\hline 3. \textbf{Case~I}. Determine the subcase based on $\mu_1^0$,
$\mu_2^0$, $\mu_\mathrm{ma}^0$, and $\lambda^0$. For  Subcase I-1,
the relay's optimal strategy is $\mathbf{B}_i =
\mathbf{V}_{\mathrm{r}i}\mathbf{P}_{\mathrm{r}i}(\lambda^0)
\mathbf{V}_{\mathrm{r}i}^\mathrm{H}$ while the source nodes solve
problem \eqref{e:Sub2} for transmission power minimization. For
Subcase I-2, use Steps~2~to~6 of the algorithm in Table~\ref{t:Al_I3} for deriving
the optimal strategies for both the source nodes and the relay. \\
\hline 4. \textbf{Case~II}. Determine the subcase based on $\mu_1^0$,
$\mu_2^0$, $\mu_\mathrm{ma}^0$, and $\lambda^0$. For Subcases II-1,
II-2, and II-3, the optimal strategy for source $i$ is $\mathbf{D}_i^0$
and the relay minimizes its transmission power via solving the
problems \eqref{e:SubII1} or \eqref{e:SubII2}. For Subcase II-4,
substitute $R^{\rm max}=\sum\limits_l \hat{R}_{\mathrm{r}l}(\lambda^0)$
in Step~3 of Table~\ref{t:Al_I3} by $R^{\rm max}= R^{\rm ma}
(\mathbf{D}^0)$ and use Steps~2~to~6 of the algorithm in Table~\ref{t:Al_I3} for
finding the optimal strategies for both the source nodes and the
relay. \\
\hline \hline
\end{tabular*}
\end{center}
\vspace{-0.5cm}
\end{table}

Concluding Case~I and Case~II, the complete procedure of deriving
the optimal solution to the problem of sum-rate maximization with
minimum transmission power for the scenario of network
optimization is summarized in Table~\ref{t:SumAl}.
\vspace{-2mm}

\subsection{Discussion: efficiency and the effect of asymmetry}
In the previous two subsections, we find solutions of the network
optimization problem in different subcases. Given the solutions
found in the previous subsections, these subcases can now be
compared and related to each other for more insights.

\textcolor{black}{
The solutions found in all subcases are \emph{optimal} in the sense that they achieve the maximum achievable sum-rate with the minimum possible power consumption. However, the optimal solutions in different subcases may not be equally good from another viewpoint which is power efficiency at the relay and the source nodes. Specifically, although the power allocation of the source nodes
and the relay jointly maximizes the sum-rate of the TWR over the
MA and BC phases at optimality, the power allocation of these
nodes may not be optimal in their individual phase of
transmission, which is MA phase for the source nodes and BC phase
for the relay. In fact, the power allocations in the two phases
have to compromise with each other in order to achieve optimality
over two phases. It is so because of the rate balancing
constraints \eqref{e:excons2} and \eqref{e:excons1}. It infers
that there is a cost of coordinating the relay and source nodes to
achieve optimality over two phases. This cost can be very
different depending on the specific subcase. In order to show the
difference in this cost, we use the metric \emph{efficiency}
defined next. A given power allocation of the relay (source nodes)
is considered as \emph{efficient} if it maximizes the BC (MA)
phase sum-rate with the actual power consumption of this power
allocation. For example, if the power allocation of the relay
consumes the power of $P_\mathrm{r} \leq
P_\mathrm{r}^\mathrm{max}$ at optimality and achieves sum-rate
$R^\mathrm{bc}$ in the BC phase, then this power allocation is
efficient if $R^\mathrm{bc}$ is the maximum achievable sum-rate in
the BC phase with power consumption $P_\mathrm{r}$. It is
inefficient otherwise. It can be shown that the chance that the
optimal power allocation is efficient for both the relay and the
source nodes is small (such situation is guaranteed to happen in
Subcase~II-1 and it is possible only in one another subcase, i.e.,
Subcase~I-1). Therefore, a joint power allocation of the relay and
source nodes is considered to be inefficient if it is inefficient
for both the relay and the source nodes, and it is considered to
be efficient otherwise. The following conclusions can be drawn for
the scenario of network optimization.
}

\textcolor{black}{
First, it can be shown that the optimal power allocation is efficient in Subcase I-1 and generally inefficient in Subcase I-2. Specifically, the optimal power allocation of the relay is always efficient in Subcase I-1 while the optimal power allocation of the source nodes can be either efficient or inefficient. In contrast, the optimal power allocation of the relay is always inefficient in Subcase I-2 while the optimal power allocation of the source nodes is also inefficient in general. For Case~II, the optimal power allocation is efficient in Subcases II-1, II-2, and II-3 and generally inefficient in Subcase II-4. Specifically, the optimal power allocation of the source nodes is efficient in Subcases II-1, II-2, and II-3 and generally inefficient in Subcase II-4 while the optimal power allocation of the relay is efficient in Subcase II-1 and inefficient in Subcases II-2, II-3, and II-4.
}

\textcolor{black}
{
Second, the optimal power allocation in Subcase I-1 achieves $\hat{R}_{\mathrm{r}1}
(\lambda^0)+\hat{R}_{\mathrm{r}2}(\lambda^0)$. In this subcase, the source nodes minimize their
power consumption while achieving the maximum sum-rate and in general they do not use up all their available power at optimality. Unlike Subcase I-1, both source nodes may use up their
available power in Subcase I-2 while the achieved sum-rate is smaller than $\hat{R}_{\mathrm{r}1}
(\lambda^0)+\hat{R}_{\mathrm{r}2}(\lambda^0)$. Similarly, the optimal
power allocation in Subcases II-1, II-2, and II-3 achieves $R^{\rm ma} (\mathbf{D}^0)$ while the relay not necessarily uses up its available power. In contrast, the optimal power allocation in Subcase II-4 consumes all the available power of the relay while the achieved sum-rate is smaller than $R^{\rm ma} (\mathbf{D}^0)$. Therefore, it can be seen that for Subcase I-1 and Subcases II-1, II-2, and II-3, in which the optimal power allocation is efficient, either the maximum possible sum-rate of the MA phase or that of the BC phase can be achieved at optimality. Moreover, the source nodes and the relay generally do not both use up their available power. In Subcases I-2 and II-4, in which the optimal power allocation is inefficient, the achieved sum-rate is however smaller than either the maximum possible sum-rate
of the MA phase or that of the BC phase, while it is possible that all nodes use up their available power.
}

Third, it can be shown for Case I that the difference between
$\max\limits_i \{1/\mu_i^0\}$ and $\min\limits_i \{1/\mu_i^0\}$
increases in general as the subcase changes from Subcase I-1 to Subcase I-2.
Similar result can be observed in Case II. As the
subcase changes from Subcase II-1, via Subcases II-2 and II-3, to
Subcase II-4, the difference between $\max\limits_i \{1/\mu_i^0\}$
and $\min\limits_i \{1/\mu_i^0\}$ increases.

Last, from the definitions of $\mu_i^0, \forall i$, it can be
seen that large difference between $\max\limits_i \{1/\mu_i^0\}$
and $\min\limits_i \{1/\mu_i^0\}$ can be, and most likely is, a
result of asymmetry in the number of antennas, available power,
and/or channel statistics at the two source nodes. It will also
be shown in detail later in the simulations that such asymmetry
can increase the occurrence of Subcases I-2 and II-4. In
contrary, if the two source nodes have same number of antennas,
same available power and same channel matrices, then
$1/\mu_1^0=1/\mu_2^0>1/\mu_{\rm ma}^0$. As a result, only Subcase
I-1 and Subcase II-1 are possible, in which the optimal power allocation is efficient. Combining this fact with the
observations in the above three paragraphs, it can be seen that the
asymmetry in the number of antennas, available power,
and/or channel statistics at the two source nodes can lead to a degradation
in the power allocation efficiency for the considered scenario of
network optimization. \textcolor{black}{As efficiency reveals the cost of coordination between the relay and source nodes required to achieve optimality over the two phases in the network optimization scenario, it can be seen that such cost is low in the case of source node symmetry and high otherwise.}

\vspace{-2mm}
\section{Simulations}\label{s:simula}
In this section, we provide simulation examples for some results
presented earlier and demonstrate the proposed algorithm for network
optimization in Table~\ref{t:Al_I3}. The general setup is as
follows. The elements of the channels $\mathbf{H}_{\mathrm{r}i}$
and $\mathbf{H}_{i\mathrm{r}}, \forall i$ are generated from
complex Gaussian distribution with zero mean and unit covariance.
The noise powers $\sigma_i^2, \forall i$ and $\sigma_\mathrm{r}^2$ are
set to 1. The rates $R^{\rm ma}(\mathbf{D})$, $\bar{R}_{i
\mathrm{r}} (\mathbf{B}_i)$, and $\hat{R}_{\mathrm{r}i}
(\mathbf{D}_{i} )$ are briefly denoted as $R^{\rm ma}$,
$\bar{R}_{i\mathrm{r}}$ and $\hat{R}_{\mathrm{r}i}$, respectively,
in all figures.

\begin{figure}[!t]
\centering \subfloat[Convergence of $R^{\rm tw}(\mathbf{B},
\mathbf{D})$, $R^{\rm ma}(\mathbf{D})$ and $R(\mathbf{B},
\mathbf{D})$]
{\includegraphics[angle=0,width=0.47\textwidth]{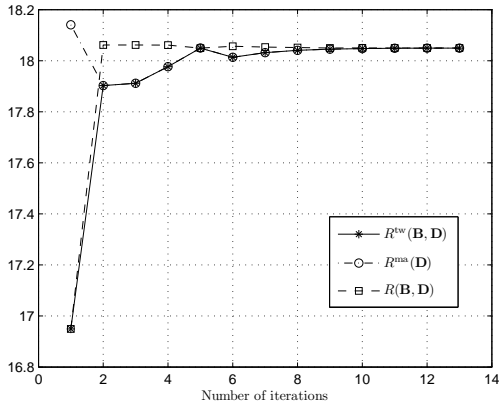}
\label{CaseI3Bfig1}}
\hspace{5mm} \subfloat[$\bar{R}_{i\mathrm{r}}(\mathbf{D}_i)$,
$\hat{R}_{\mathrm{r}i}(\lambda_i), \forall i$ and the power
consumption of the source nodes during the iterations]
{\includegraphics[angle=0,width=0.47\textwidth]{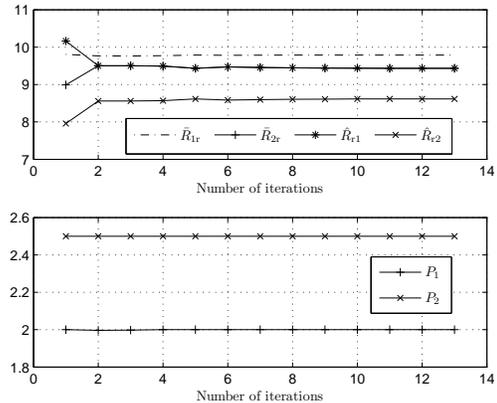}
\label{CaseI3Bfig2}}
\caption{Illustration of the algorithm in
Table~\ref{t:Al_I3}.}\label{CaseI3Bfig}
\vspace{-7mm}
\end{figure}

\emph{Example 1: The process of finding the optimal solution for
network optimization Subcase I-2 using the proposed algorithm in
Table~\ref{t:Al_I3}.} The specific setup for this example is as
follows. The number of antennas $n_1, n_2$, and $n_\mathrm{r}$ are
set to be $6, 4$ and $8$, respectively. Power limits for the
source nodes are $P_1^{\mathrm{max}}=2, P_2^{\mathrm{max}}=2.5$.
The relay's power limit is set to $P_r^{\mathrm{max}}=3$. Since
the optimality of the solution derived using the algorithm has
been proved analytically by Theorem~3, we focus on demonstrating
the iterative process and the convergence of the algorithm.
Fig.~\ref{CaseI3Bfig1} shows instantaneous $R^{\rm tw}(\mathbf{B},
\mathbf{D})$, $R^{\rm ma}(\mathbf{D})$ and $R(\mathbf{B},
\mathbf{D})$ versus the number of iterations. From the figure, it
can be seen that the above three rates converge very fast.
Fig.~\ref{CaseI3Bfig2} shows the instantaneous
$\bar{R}_{i\mathrm{r}} (\mathbf{D}_i)$, $\hat{R}_{\mathrm{r}i}
(\lambda_i), \forall i$ and the power consumption of the source
nodes~1~and~2, denoted as $P_1$ and $P_2$, respectively. Two
observations can be drawn from Fig.~\ref{CaseI3Bfig2}. First,
$\hat{R}_{\mathrm{r}2}(\lambda_2)<\bar{R}_{1\mathrm{r}}(\mathbf{D}_1)$
and $\hat{R}_{\mathrm{r}1}(\lambda_1) = \bar{R}_{2\mathrm{r}}
(\mathbf{D}_2)$ in the optimal solution since the sum-rate is
bounded by $R^{\rm ma}(\mathbf{D})<\bar{R}_{1\mathrm{r}}
(\mathbf{D}_1) + \bar{R}_{2\mathrm{r}}(\mathbf{D}_2)$. Second,
both source nodes use all available power in the optimal solution.
The latter observation verifies the conclusion that for Case~I the optimal
power allocation in Subcase I-2 is inefficient for using possibly more power and achieving less
sum-rate.

\begin{figure}[!t]
\centering \subfloat[Percentage of the increase in sum-rate at optimality]
{\includegraphics[angle=0,width=0.47\textwidth]{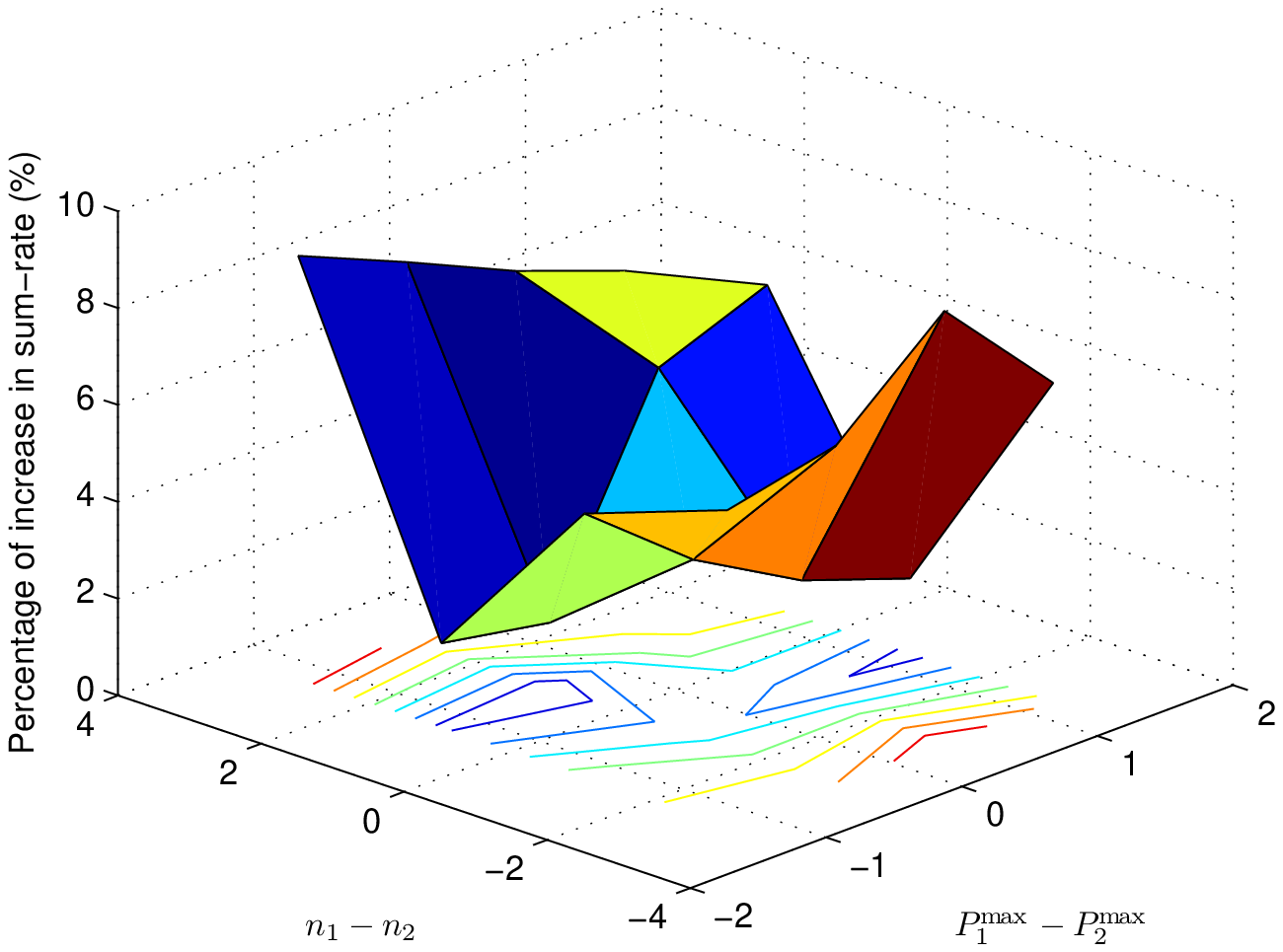}
\label{f:Comp_R}}
\hspace{5mm} \subfloat[percentage of the decrease in power consumption at optimality]
{\includegraphics[angle=0,width=0.47\textwidth]{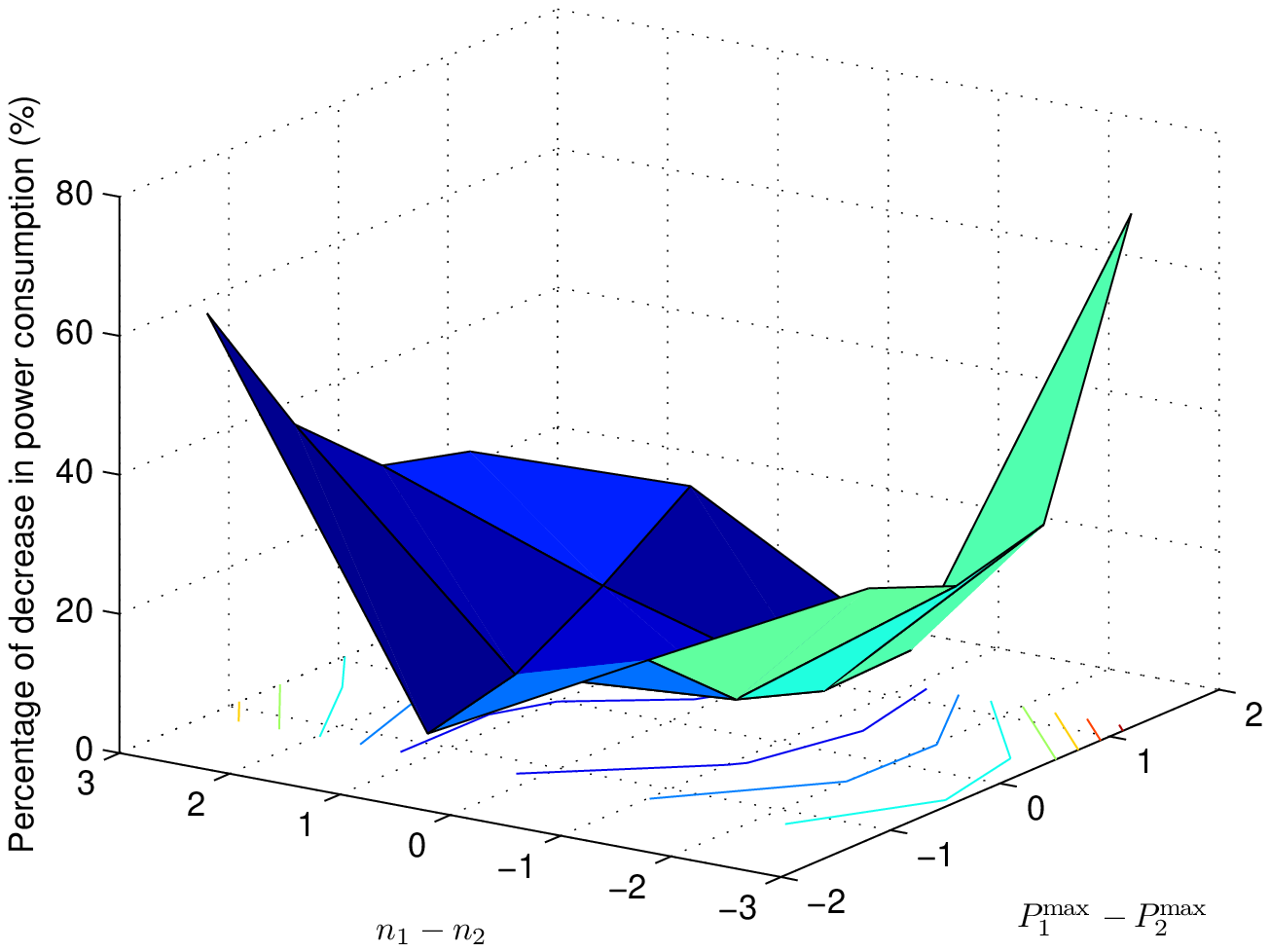}
\label{f:Comp_P}}
\caption{Improvements as compared to relay optimization}
\vspace{-4mm}
\end{figure}

\emph{Example 2: Comparison with relay optimization in Part~I}. The specific setup for this example is as follows. The number of antennas at the relay, i.e., $n_\mathrm{r}$, is set to be $5$. The power limit of the relay, i.e., $P^{\rm max}_\mathrm{r}$ is set to be 3. The total number of antennas at both source nodes is fixed so that $n_1+n_2=5$. The total available power at both source nodes is also fixed so that $P^{\rm max}_1+P^{\rm max}_2=2$. Given the above total number of antennas and total available power at the source nodes, both the relay optimization and the network optimization problems are solved for different $n_1$, $n_2$, $P^{\rm max}_1$, and $P^{\rm max}_2$ for 100 channel realizations. The percentage of the increase in the average sum-rate and the percentage of the decrease in the average power consumption at optimality of the network optimization problem compared to those at optimality of the relay optimization problem are plotted in Figs.~\ref{f:Comp_R}~and~\ref{f:Comp_P}, respectively. These percentages are shown versus the difference between the number of antennas and the difference between the power limits at the source nodes. From these two figures, it can be seen that although the optimal solution of the network optimization problem on average consumes much less power than that of the relay optimization problem, it still achieves larger sum-rate.
%Note that the relatively small increase in sum-rate is because we use power allocation which maximizes the MAC phase sum-rate for the source nodes in the relay optimization scenario.
Moreover, it can also be seen that the improvements, in either sum-rate or power consumption of the optimal solution of the network optimization problem as compared to that of the relay optimization problem, become more obvious when there is more asymmetry in the system. This is because the source nodes and the relay can jointly optimize their power allocations and therefore cope with (to some extent) the negative effect of the asymmetry in the system in the network optimization scenario. In contrast, the relay optimization scenario does not has such capability to combat the negative effect of asymmetry.

\begin{figure}[!t]
\begin{center}
\includegraphics[angle=0,width=0.50\textwidth]{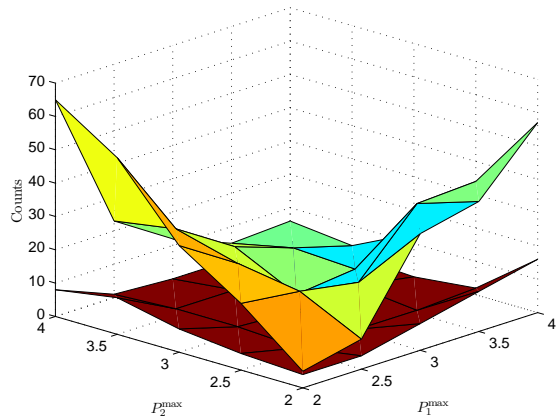}
\end{center}
\vspace{-5mm}
\caption {The number of times that Subcases I-2 and II-4 appear with asymmetry in source nodes' power limits.} \label{f:asymP}
\vspace{-5mm}
\end{figure}

\emph{Example 3: The effect of asymmetry in the scenario of
network optimization.} First, we solve the network optimization
problem for different $P_1^{\mathrm{max}}$ and
$P_2^{\mathrm{max}}$ given that $P_\mathrm{r}^{\mathrm{max}}$ is
fixed. The number of antennas of the relay is set to 8 and the
number of antennas of both source nodes is set to 4. For each
combination of $P_1^{\mathrm{max}}$ and $P_2^{\mathrm{max}}$, we
use 200 channel realizations and solve the resulting 200 network
optimization problems. The number of times that
Subcases~I-2~and~II-4 appear are plotted in Fig.~\ref{f:asymP}.
In this figure, the points in the upper surface correspond to the
counts of Subcase I-2 while the points in the lower surface
correspond to the counts of Subcase II-4. From
Fig.~\ref{f:asymP}, it can be seen that in general the count of
either Subcase I-2 or Subcase II-4 is the smallest when
$P_1^{\mathrm{max}}=P_2^{\mathrm{max}}$. Moreover, for any given
$P_1^{\mathrm{max}}$ or $P_2^{\mathrm{max}}$, the largest count of
either Subcase I-2 or Subcase II-4 mostly happens where the
difference between $P_1^{\mathrm{max}}$ and $P_2^{\mathrm{max}}$
is the largest.\footnote{Note, however, that subcases are also determined
by the ratio of the number of antennas at the relay to the number
of antennas at the source nodes, the ratio of
$P_\mathrm{r}^{\mathrm{max}}$ to $P_i^{\mathrm{max}}, \forall i$,
the channel realizations and other factors, instead of only by
$P_i^{\mathrm{max}}, \forall i$.} The above two observations are
accurate for most of the times in Fig.~\ref{f:asymP}, which
shows that the asymmetry of $P_i^{\mathrm{max}}$ leads to
the rise of the occurrence of Subcases I-2 and II-4.

\begin{figure}[!t]
\centering \subfloat[Sum of the Counts of Subcases I-2 and II-4
versus $P_1^{\mathrm{max}}$ and $P_2^{\mathrm{max}}$, $n_1=4, n_2=6$]
{\includegraphics[angle=0,width=0.45\textwidth]{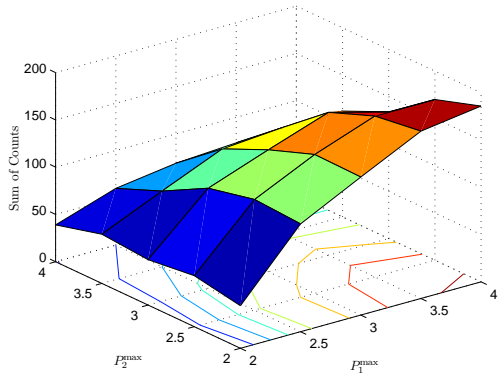}
\label{f:asymM1}}
\hspace{5mm} \subfloat[Sum of the Counts of Subcases I-2 and II-4
versus $P_1^{\mathrm{max}}$ and $P_2^{\mathrm{max}}$, $n_1=n_2=6$]
{\includegraphics[angle=0,width=0.45\textwidth]{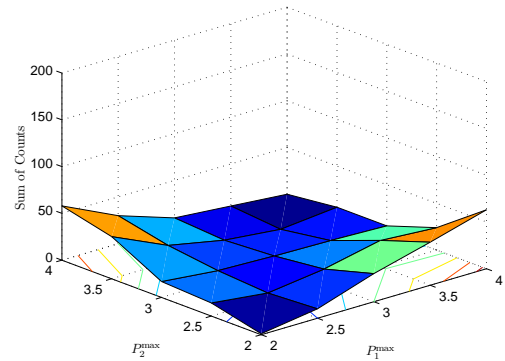}
\label{f:asymM2}}
\vspace{-1mm} \caption{Illustration of the effect of asymmetry in
the number of antennas at the source nodes.}\label{f:asymM}
\vspace{-5mm}
\end{figure}

Next we demonstrate the effect of asymmetry in the number of
antennas at the source nodes. The number of antennas of the relay
is still 8 and $P_\mathrm{r}^{\mathrm{max}}$ is still 4. However,
the number of antennas of sources nodes 1~and~2 are first set to 4
and 6 and then 6 both, respectively. The network optimization
problem is solved for different $P_1^{\mathrm{max}}$ and
$P_2^{\mathrm{max}}$ and the sum of the counts of
Subcases~I-2~and~II-4 in 200 channel realizations is plotted in
Fig.~\ref{f:asymM} for each combination of $P_1^{\mathrm{max}}$
and $P_2^{\mathrm{max}}$. From Fig.~\ref{f:asymM1}, it can be seen
that the sum of the counts of Subcases~I-2~and~II-4 substantially
increases when $n_1=4$ and $n_2=6$ as compared to the sum of the
counts in Fig.~\ref{f:asymP} on most of the points. However, as
shown in Fig.~\ref{f:asymM2}, when $n_1=n_2=6$, the sum of the counts of
Subcases~I-2~and~II-4 drops to the same level as the sum of the
counts in Fig.~\ref{f:asymP}. Therefore, it can be seen that asymmetry in the number of
antennas at the source nodes leads to larger chance of
Subcases~I-2~and~II-4.

\begin{figure}[!t]
\centering \subfloat[Sum of the Counts of Subcases I-2 and II-4
versus $v_1$ and $v_2$ (without assuming channel reciprocity)]
{\includegraphics[angle=0,width=0.45\textwidth]{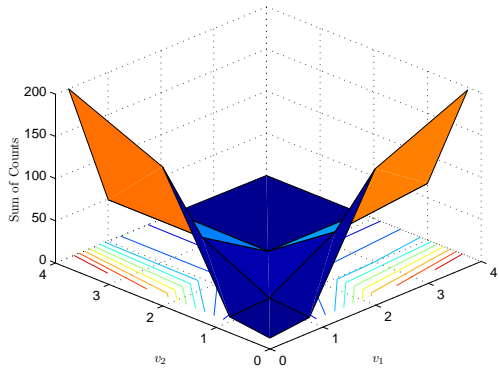}
\label{f:asymH1}}
\hspace{5mm} \subfloat[Sum of the Counts of Subcases I-2 and II-4
versus $v_1$ and $v_2$ (assuming channel reciprocity)]
{\includegraphics[angle=0,width=0.45\textwidth]{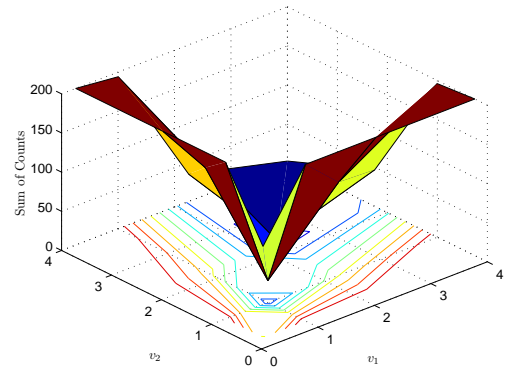}
\label{f:asymH2}}
\caption{Illustration of the effect of asymmetry in channel
statistics.}\label{f:asymH} \vspace{-5mm}
\end{figure}

Lastly, we show the effect of asymmetry in channel statistics.
Instead of generating the real and imaginary parts of each element
of $\mathbf{H}_{i\mathrm{r}}, \forall i$ from Gaussian
distributions with zero mean and unit variance, here we use
Gaussian distribution with zero mean and variance $v_{i}$ to
generate the real and imaginary parts of each element of
$\mathbf{H}_{i\mathrm{r}}, \forall i$. For each combination of
$v_1$ and $v_2$, we use 200 channel realizations and solve the
resulting 200 network optimization problems. The number of
antennas at the relay is set to 6 and the number of antennas at
both source nodes is set to 4. The power limits are
$P_\mathrm{r}^{ \mathrm{max}} =5$ and $P_i^{\mathrm{max}}=3,
\forall i$. The sum of the counts of Subcases~I-2~and~II-4 is
plotted in Fig.~\ref{f:asymH} versus $v_1$ and $v_2$.
Fig.~\ref{f:asymH1} corresponds to the case without assuming
channel reciprocity, in which the real and imaginary parts of each
element of $\mathbf{H}_{\mathrm{r}i}, \forall i$ are generated
from Gaussian distributions with zero mean and unit variance.
Fig~\ref{f:asymH2} corresponds to the case of reciprocal channels,
i.e., $\mathbf{H}_{\mathrm{r}i} = \mathbf{H}_{i
\mathrm{r}}^\mathrm{T}, \forall i$ where $(\cdot)^\mathrm{T}$
represents transpose. It can be seen from both Fig.~\ref{f:asymH1}
and Fig.~\ref{f:asymH2} that the sum of the counts of Subcases
I-2~and~II-4 tends to increase when the difference between $v_1$
and $v_2$ becomes larger. Therefore, Fig.~\ref{f:asymH} clearly
shows that the asymmetry in the channel statistics also leads to
larger chance of Subcases~I-2~and~II-4.

\section{Conclusion}\label{s:conclu}
In Part II of this two-part paper, we have solved the problem of sum-rate maximization using minimum transmission power for MIMO DF TWR in the scenario of network optimization. For finding the optimal solution, we study the original problem in two cases each of which has several subcases. It has been shown that for all except two subcases, the original problem can be simplified into corresponding convex optimization problems. For the remaining two subcases, we have found the properties that the optimal solution must satisfy and have proposed the algorithm to find the optimal solution based on these properties. We have shown that the optimal power allocation in these two subcases are inefficient in the sense that it always consumes all the available power of the relay (and sometimes all the available power of the source nodes as well) yet cannot achieve the maximum sum-rate of either the MA or BC phase. We have also shown that the asymmetry in the number of antennas, power limits, and channel statistics leads to a higher probability of the above-mentioned two subcases. Combining with Part~I of this work, we have provided a complete and detailed study of sum-rate maximization using minimum power consumption for MIMO DF TWR.

\section{Appendix}\label{s:appen}

\subsection{Proof of Lemma~1}\label{s:PLm1II}

\textcolor{black}{
Proof for claim 1: Given $\widetilde{\mathbf{D}}$ as defined in the lemma, it follows that $R^{\rm ma}(\widetilde{\mathbf{D}})=\bar{R}_{j\rm r}(\widetilde{\mathbf{D}}_j)$. From the definitions \eqref{e:watermu1}-\eqref{e:watermum}, it can be seen that $R^{\rm ma}(\widetilde{\mathbf{D}})>\bar{R}_{j\rm r}(\widetilde{\mathbf{D}}_j) = \bar{R}_{j\rm r}(\mathbf{D}_j)$ if $1/\mu_{\rm ma}(\widetilde{\mathbf{D}})> 1/\mu_j$. Therefore, it is necessary that $1/\mu_{\rm ma}(\widetilde{\mathbf{D}})\leq 1/\mu_j$.
}

\textcolor{black}{
Proof for claim 2: First, note that $R^{\rm ma}(\widehat{\mathbf{D}})$ is a continuous and strictly increasing function of $t$ in $[0, 1]$. Second, based on the definition \eqref{e:watermum}, it follows that $R^{\rm ma}(\widehat{\mathbf{D}})$ is a strictly increasing function of $1/\mu_{\rm ma}(\widehat{\mathbf{D}})$ when $1/\mu_{\rm ma}(\widehat{\mathbf{D}})> \min \{1/\alpha_i(k), \forall i, \forall k\}$, or equivalently, $R^{\rm ma}(\widehat{\mathbf{D}})>0$. Since $\text{Tr}\{\mathbf{D}_1\}>0$ and $\text{Tr}\{\mathbf{D}_2\}>0$, we have $R^{\rm ma}(\widehat{\mathbf{D}})>0$ for all $t\in[0, 1]$. Thus, given the fact that $1/\mu_{\rm ma}(\widetilde{\mathbf{D}})\!\leq 1\!/\mu_j$ when $t\!=\!0$ and that $1/\mu_{\rm ma}(\widetilde{\mathbf{D}})= 1/\mu_{\rm ma}(\mathbf{D})> 1/\mu_j$ when $t=1$, it can be seen that there exists $\widehat{t}\in [0, 1)$ such that $1/\mu_{\rm ma}(\widehat{\mathbf{D}}) = 1/\mu_j$ when $t=\widehat{t}$. Using Lemma~1 in Part~I of this two-part paper \cite{PartI}, i.e., $1/\mu_{\rm ma}\!<\!\max\{1/\mu_1, 1/\mu_2\}$, it can be seen that $1/\mu_i(\widehat{\mathbf{D}}_i)> 1/\mu_{\rm ma}(\widehat{\mathbf{D}}) = 1/\mu_j$ when $t=\widehat{t}$. \hfill$\blacksquare$
}

\vspace{-1mm}
\subsection{Proof of Lemma~2}\label{s:PLm2II}

Consider the first part of constraint \eqref{e:propty1}, i.e., $\lambda_j=\mu_i>\mu_{\rm ma}$ if $\lambda_i<\lambda_j$. Using Lemma~2 in Part~I \cite{PartI}, it can be seen that $\lambda_i, \forall i$ satisfy $\lambda_1 =
\lambda_2$ if $\min\{1/\mu_i\}\geq 1/\mu_\mathrm{ma}$ at optimality. Therefore, we have $\min\{1/\mu_i\}< 1/\mu_{\rm ma}$
given that $\lambda_1 \neq \lambda_2$. Using the same lemma and the constraint \eqref{e:WLcons1NT}, it can be further concluded that $1/\mu_i< 1/\mu_{\rm ma}$ at optimality given that
$\lambda_i<\lambda_j$. Otherwise, the constraint \eqref{e:WLcons2NT}
cannot be satisfied. Therefore, $1/\mu_j > 1/\mu_{\rm ma}$
according to Lemma~1 in Part~I \cite{PartI}. Due to the constraint \eqref{e:WLcons1NT},  we
must have $1/\lambda_j\leq 1/\mu_i$ at optimality. Moreover,
from Lemma~2 in Part~I and the assumption that $\lambda_i<\lambda_j$, it can
be seen that $1/\lambda_j< 1/\mu_i$ is not optimal. Therefore,
$1/\lambda_j=1/\mu_i$ if $\lambda_i<\lambda_j$. Following the same approach, the second part, i.e., $\lambda_j=\mu_i>\mu_{\rm ma}$ if $\mu_i>\mu_{\rm ma}$ can be proved similarly.

\vspace{-1mm}
\subsection{Proof of Theorem~1}\label{s:PTh1}

\textcolor{black}{
Recall the definitions of $\mu_1$, $\mu_2$, and $\mu_{\rm ma}$ in \eqref{e:watermu1}-\eqref{e:watermum}. Considering the constraints \eqref{e:Sub2cons1}-\eqref{e:Sub2cons3} in the problem \eqref{e:Sub2}, it can be seen that at optimality we must have $\mu_{\rm ma}^\star\leq \lambda^0$, $\mu_{1}^\star\leq \lambda^0$, and $\mu_{2}^\star\leq \lambda^0$. Otherwise, the above mentioned constraints cannot be satisfied. We will prove Theorem~1 by contradiction.
}

\textcolor{black}{
Assume that $\mu_{\rm ma}^\star\neq \lambda^0$ at optimality, then $\mu_{\rm ma}^\star < \lambda^0$ according to the above paragraph. Using Lemma~1 in Part~I of this two-part paper \cite{PartI}, i.e, $1/\mu_{\rm ma}<\max\{1/\mu_1, 1/\mu_2\}$, and given that $\mu_{1}^\star\leq \lambda^0$ and $\mu_{2}^\star\leq \lambda^0$, there are only two possible cases as follows: a) $\max\{1/\mu_1^\star, 1/\mu_2^\star\}> 1/\mu_{\rm ma}^\star > \min\{1/\mu_1^\star, 1/\mu_2^\star\}\geq 1/\lambda^0$ and b) $\max\{1/\mu_1^\star, 1/\mu_2^\star\}\geq \min\{1/\mu_1^\star, 1/\mu_2^\star\}\geq 1/\mu_{\rm ma}^\star> 1/\lambda^0$. Assume without loss of generality that $\max\{1/\mu_1^\star, 1/\mu_2^\star\}=1/\mu_1^\star$ and $\min\{1/\mu_1^\star, 1/\mu_2^\star\}=1/\mu_2^\star$. If it is Case a), then we have $1/\mu_1^\star > 1/\mu_{\rm ma}^\star > 1/\mu_2^\star\geq 1/\lambda^0$. Use Lemma~1 (of Part II) with $\widehat{\mathbf{D}}_i=t\mathbf{D}_1^\star$ and $\widehat{\mathbf{D}}_j=\mathbf{D}_2^\star$. As proved in Lemma~1, there exists $t \in [0,1)$ such that $\mu_1(t\mathbf{D}_1^\star)>1/\mu_{\rm ma}([t\mathbf{D}_1^\star, \mathbf{D}_2^\star])= 1/\mu_2^\star$. Since $1/\mu_2^\star \geq 1/\lambda^0$, we have $\mu_1(t\mathbf{D}_1^\star)>1/\mu_{\rm ma}([t\mathbf{D}_1^\star, \mathbf{D}_2^\star])= 1/\mu_2^\star \geq 1/\lambda^0$, which indicates that $\hat{\mathbf{D}}=[t\mathbf{D}_1^\star, \mathbf{D}_2^\star]$ also satisfies \eqref{e:Sub2cons1}-\eqref{e:Sub2cons4} while $\text{Tr}\{t\mathbf{D}_1^\star\} + \text{Tr}\{\mathbf{D}_2^\star\} < \text{Tr}\{\mathbf{D}_1^\star\} + \text{Tr}\{\mathbf{D}_2^\star\}$. It contradicts the fact that $\mathbf{D}^\star = [\mathbf{D}_1^\star, \mathbf{D}_2^\star]$ is the optimal solution to the problem \eqref{e:Sub2}. Therefore, Case a) is impossible. If it is Case b), there exist two following possible subcases: subcase b-1) there exists $i\in\{1,2\}$ such that $1/\mu_{\rm ma}(\widehat{\mathbf{D}})=1/\lambda^0$ and $1/\mu_i(\widehat{\mathbf{D}}_i) \geq 1/\mu_{\rm ma}(\widehat{\mathbf{D}})$ where $\widehat{\mathbf{D}}=[\widehat{\mathbf{D}}_1, \widehat{\mathbf{D}}_2]$ with $\widehat{\mathbf{D}}_i=t_i\mathbf{D}_i^\star$ and $\widehat{\mathbf{D}}_j=\mathbf{D}_j^\star$ for some $t_i\in [0,1)$ and subcase b-2) there does not exist $t_i\in [0, 1)$ such that $1/\mu_{\rm ma}(\widehat{\mathbf{D}})=1/\lambda^0$ and $1/\mu_i(\widehat{\mathbf{D}}_i) \geq 1/\mu_{\rm ma}(\widehat{\mathbf{D}})$ where $\widehat{\mathbf{D}}=[\widehat{\mathbf{D}}_1, \widehat{\mathbf{D}}_2]$ with $\widehat{\mathbf{D}}_i=t_i\mathbf{D}_i^\star$ and $\widehat{\mathbf{D}}_j=\mathbf{D}_j^\star$ for either $i=1$ or $i=2$. In subcase b-1), it can be seen that $\widehat{\mathbf{D}}$ satisfies \eqref{e:Sub2cons1}-\eqref{e:Sub2cons4} while $\text{Tr}\{t_i\mathbf{D}_i^\star\} + \text{Tr}\{\mathbf{D}_j^\star\} < \text{Tr}\{\mathbf{D}_1^\star\} + \text{Tr}\{\mathbf{D}_2^\star\}$. It contradicts the fact that $\mathbf{D}^\star = [\mathbf{D}_1^\star, \mathbf{D}_2^\star]$ is the optimal solution to the problem \eqref{e:Sub2}. Therefore, subcase b-1) is impossible. If it is subcase b-2), it indicates that with $t_i\in [0, 1)$, for either $i=1$ or $i=2$, such that $1/\mu_{\rm ma}(\widehat{\mathbf{D}})=1/\lambda^0$, we have $1/\mu_i(\widehat{\mathbf{D}}_i)=1/\mu_i(t_i\widehat{\mathbf{D}}_i^\star)<1/\mu_{\rm ma}(\widehat{\mathbf{D}})=1/\lambda^0$. As a result, there exists $t_i^\prime\in(t_i, 1)$ such that $1/\mu_i(t_i^\prime\mathbf{D}_i^\star)=1/\lambda^0$ and $1/\mu_{\rm ma}(\mathbf{D}^\prime)> 1/\lambda^0$ where $\mathbf{D}^\prime=[\mathbf{D}_1^\prime, \mathbf{D}_2^\prime]$ with $\mathbf{D}_i^\prime=t_i^\prime\mathbf{D}_i^\star$ and $\mathbf{D}_j^\prime=\mathbf{D}_j^\star$. Note that $1/\mu_{\rm ma}(\mathbf{D}^\prime)> 1/\lambda^0$ because if $1/\mu_i(\mathbf{D}_i^\prime)=1/\lambda^0$ and $1/\mu_{\rm ma}(\mathbf{D}^\prime)= 1/\lambda^0$ then it is subcase b-1) instead of subcase b-2). Recalling that $1/\mu_j(\mathbf{D}_j^\star)>1/\mu_{\rm ma}(\mathbf{D}^\star)>1/\mu_{\rm ma}(\mathbf{D}^\prime)$, we have $1/\mu_j(\mathbf{D}_j^\star)>1/\mu_{\rm ma}(\mathbf{D}^\prime)>1/\mu_i(\mathbf{D}_i^\prime)=1/\lambda_0$. It indicates that by changing $\mathbf{D}_i^\star$ in the optimal solution to $\mathbf{D}_i^\prime=t_i^\prime\mathbf{D}_i^\star$ (and thus using less power than $\text{Tr}\{\mathbf{D}_1^\star\} + \text{Tr}\{\mathbf{D}_2^\star\}$ while satisfying \eqref{e:Sub2cons1}-\eqref{e:Sub2cons4}), subcase b-2) changes to Case a). As it is proved that Case a) is impossible at optimality, so it is subcase b-2). %for leading to less power consumption than $\text{Tr}\{\mathbf{D}_1^\star\} + \text{Tr}\{\mathbf{D}_2^\star\}$ while satisfying \eqref{e:Sub2cons1}-\eqref{e:Sub2cons4} , so it is subcase b-2).
}

\textcolor{black}{
Therefore, it is proved that the assumption $\mu_{\rm ma}^\star\neq \lambda^0$ must lead to either of two cases while both of them are impossible at optimality. Thus, it is impossible that $\mu_{\rm ma}^\star\neq \lambda^0$. As a result, we must have  $\mu_{\rm ma}^\star= \lambda^0$. This completes the proof. \hfill $\blacksquare$
}

\vspace{-1mm}
\subsection{Proof of Theorem~2}\label{s:PTh2}
Proof of property 1. First we show that  $1/\mu_\mathrm{ma}^* <
1/\lambda^0$. Since the maximum $\bar{R}_{j\mathrm{r}}
(\mathbf{D}_j)$, as the objective function of the problem \eqref{e:Sub3}, cannot achieve $\hat{R}_{\mathrm{r}i}(\lambda^0)$
in Subcase I-2, it can be seen that $1/\mu_j<
1/\lambda^0$ whenever $1/\mu_\mathrm{ma} \geq 1/\lambda^0$ and
$1/\mu_i \geq 1/\lambda^0$. As a result, any $\mathbf{D}$ such
that $1/\mu_\mathrm{ma}\geq 1/\lambda^0$ is not optimal. The
reason is that in such a case the optimal relay power allocation
requires $1/\lambda_i=1/\mu_j < 1/\lambda^0$ according to Lemma~2
and such relay power allocation leads to a BC phase sum-rate
$\sum\limits_l \hat{R}_{\mathrm{r}l}(\lambda_l)$ which is less
than $\hat{R}_{\mathrm{r}1}(\lambda^0) + \hat{R}_{\mathrm{r}2}
(\lambda^0)$ according to Lemma~2 in Part~I \cite{PartI}. Since $1/\mu_\mathrm{ma}\geq
1/\lambda^0$ implies that $R^{\rm ma}(\mathbf{D})\geq
\hat{R}_{\mathrm{r}1}(\lambda^0) + \hat{R}_{\mathrm{r}2}
(\lambda^0)$,  it can be seen that the constraint
\eqref{e:excons1} is not satisfied and therefore such strategies
cannot be optimal. Next we show that $\min\limits_{l}\{1/\mu_l^*\}
< 1/\mu_\mathrm{ma}^*$. Assuming that $1/\mu_\mathrm{ma}^*
\leq\min\limits_{l} \{1/\mu_l^*\}$, it leads to $1/
\mu_\mathrm{ma}^* < 1/\lambda^0$ given that the problem
\eqref{e:Sub2} is infeasible. Moreover, it also leads to the
result that $\lambda_l^* =\mu_\mathrm{ma}^*, \forall l$. However,
it is not difficult to see that $R^{\rm ma}(\mathbf{D})$,
$\sum\limits_l \hat{R}_{\mathrm{r}l}(\lambda_l)$ and eventually
$R^{\rm tw}(\mathbf{B}, \mathbf{D})$ can be increased in this case
through appropriately increasing $1/\mu_\mathrm{ma}$, which is
feasible since $1/\mu_\mathrm{ma}^0 > 1/\lambda^0> 1/
\mu_\mathrm{ma}^*$, and at least one of $1/\lambda_i$ and
$1/\lambda_j$, which is also feasible since $1/ \lambda_i^* =
1/\mu_\mathrm{ma}^*<1/\lambda^0$, given that $1/ \mu_\mathrm{ma}^*
\leq \min\limits_{l}\{1/\mu_l^*\}$ and $1/ \mu_\mathrm{ma}^* < 1 /
\lambda^0$. It contradicts the assumption that $\mathbf{D}^{*}$
and $\mathbf{B}^*$ are the optimal solution. Therefore,
$1/\mu_\mathrm{ma}^* >\min\limits_{l}\{1/\mu_l^*\}$.

Proof of property 2. Given the fact that $1/\mu_\mathrm{ma}^* <
1/\lambda^0$, the problem boils down to finding the maximum
$1/\mu_\mathrm{ma}$ such that the corresponding rate $R^{\rm
ma}(\mathbf{D})$ can also be achieved by the BC phase sum-rate
$\sum\limits_l\hat{R}_{\mathrm{r}l}(\lambda_l)$ subject to the
first constraint in \eqref{e:SumR} and the constraint that
$\min\limits_l \{1/\lambda_l\}=\min\limits_l \{1/\mu_l\}$ as
stated in Lemma~2. Since the maximum $\sum\limits_l
\hat{R}_{\mathrm{r}l} (\lambda_l)$ cannot achieve $R^{\rm
ma}(\mathbf{D})$ subject to the above-mentioned constraints as
long as $1/\mu_\mathrm{ma} \geq 1/\lambda^0$, the problem is
equivalent to finding the maximum achievable $\sum\limits_l
\hat{R}_{\mathrm{r}l}(\lambda_l)$ subject to the constraints that
$\min\limits_l \{1/\lambda_l\}=\min\limits_l \{1/\mu_l\}$ and that
$R^{\rm ma}(\mathbf{D})=\sum\limits_l \hat{R}_{\mathrm{r}l}
(\lambda_l)$. Since $R^{\rm ma}(\mathbf{D})$ can achieve up to
$R^{\rm ma}(\mathbf{D}^0)$, it is not difficult to see that the
maximum achievable $\sum\limits_l\hat{R}_{\mathrm{r}l}(\lambda_l)$
subject to the above-mentioned constraints demands the relay to
use full transmission power $P_{\rm r}^\mathrm{max}$.

Proof of property 3. Define the index $i^-$ such that
$\min\limits_l \{1/\mu_l\}=1/\mu_{i^-}$. Recall from the proof of
property 1 that $1/\mu_\mathrm{ma}^*<1/\lambda^0$. As a result,
$R^{\rm ma}(\mathbf{D}^*)$ is not the maximum $R^{\rm
ma}(\mathbf{D})$ that can be achieved, which implies that there
exists $\mathbf{D}^{\rm s}$ such that $R^{\rm ma}(\mathbf{D}^{\rm
s})>R^{\rm ma}(\mathbf{D}^*)$ and $\bar{R}_{i^-\mathrm{r}}
(\mathbf{D}_{i^-}^{\rm s}) > \bar{R}_{i^-\mathrm{r}}
(\mathbf{D}_{i^-}^*) -\delta$ where $\delta$ is a positive number.
Define $\mathbf{Z}=\mathbf{H}_{1\mathrm{r}}\mathbf{D}_1
\mathbf{H}_{1\mathrm{r}}^\mathrm{H} \!+\!\mathbf{H}_{2
\mathrm{r}}\mathbf{D}_2\mathbf{H}_{2\mathrm{r}}^\mathrm{H}$. It can be seen that $R^{\rm ma}(\mathbf{D})$ is a concave
function of $\mathbf{Z}$. If $\mathbf{D}^*$ is not the optimal
solution to the problem of maximizing $\min\limits_l \{1/\mu_l\}$
subject to the constraints in \eqref{e:ProfTh3cons}, there exists
$\mathbf{D}^{\rm q}$ such that $R^{\rm ma}(\mathbf{D}^{\rm q})\geq
R^{\rm ma}(\mathbf{D}^*)$ and $\bar{R}_{i^-\mathrm{r}}
(\mathbf{D}_{i^-}^{\rm q}) > \bar{R}_{i^-\mathrm{r}}
(\mathbf{D}_{i^-}^*)$. Then, for any $0<\alpha<1$, these exists
$\mathbf{D}^{\rm c}$ such that $\mathbf{D}^{\rm c}_l =
\alpha\mathbf{D}_l^{\rm q} + (1-\alpha)\mathbf{D}_l^{\rm s},
\forall l$. Moreover, for any $\alpha$ such that
\begin{equation}
\frac{ \bar{R}_{i^-\mathrm{r}} (\mathbf{D}_{i^-}^*) -
\bar{R}_{i^-\mathrm{r}}(\mathbf{D}_{i^-}^{\rm
s})}{\bar{R}_{i^-\mathrm{r}}(\mathbf{D}_{i^-}^{\rm q})-
\bar{R}_{i^-\mathrm{r}}(\mathbf{D}_{i^-}^{\rm s})}<\alpha<1,
\end{equation}
it can be shown that
$\bar{R}_{i^-\mathrm{r}}(\mathbf{D}_{i^-}^{\rm c}) >
\bar{R}_{i^-\mathrm{r}}(\mathbf{D}_{i^-}^*)$ using the fact that
$\bar{R}_{l\mathrm{r}}(\mathbf{D}_l)$ is concave with respect to
$\mathbf{D}_l, \forall l$. Denoting $\mathbf{\mathbf{Z}}^{\rm q} =
\mathbf{H}_{1\mathrm{r}}\mathbf{D}_1^{\rm q}
\mathbf{H}_{1\mathrm{r}}^\mathrm{H} \!+\! \mathbf{H}_{ 2
\mathrm{r}} \mathbf{D}_2^{\rm q}\mathbf{H}_{2
\mathrm{r}}^\mathrm{H}$ and $\mathbf{Z}^{\rm s}=\mathbf{H}_{1
\mathrm{r}}\mathbf{D}_1^{\rm s}\mathbf{H}_{1\mathrm{r}}^\mathrm{H}
\!+\! \mathbf{H}_{2\mathrm{r}}\mathbf{D}_2^{\rm
s}\mathbf{H}_{2\mathrm{r}}^\mathrm{H}$, it can be shown that
$\mathbf{D}_l^{\rm c}, \forall l$ lead to $\mathbf{Z}^{\rm
c}=\alpha \mathbf{Z}^{\rm q}+(1-\alpha)\mathbf{Z}^{\rm s}$ and
therefore $R^{\rm ma}(\mathbf{D}^{\rm c})\geq \alpha R^{\rm
ma}(\mathbf{D}^{\rm q})+(1-\alpha) R^{\rm ma}(\mathbf{D}^{\rm
s})>R^{\rm ma}(\mathbf{D}^*)$. Therefore, if $\mathbf{D}^*$ does
not maximize $\bar{R}_{i^-\mathrm{r}}(\mathbf{D}_{i^-})$ subject
to the constraints in \eqref{e:ProfTh3cons}, then
$\bar{R}_{i^-\mathrm{r}}(\mathbf{D}_{i^-})$ and $R^{\rm
ma}(\mathbf{D})$ can be simultaneously increased. The fact that
$\bar{R}_{i^-\mathrm{r}}(\mathbf{D}_{i^-})$ can be increased means
that $\min\limits_l\{1/\mu_l\}$ can be increased, which implies
that the BC phase sum-rate $\sum\limits_l\hat{R}_{\mathrm{r}l}
(\lambda_l)$ can be increased according to Lemma~2 in Part~I \cite{PartI} subject to the
constraint that $\min\limits_l \{1/\lambda_l\}=\min\limits_l
\{1/\mu_l\}$ as stated in Lemma~2. Given this result, the fact
that $R^{\rm ma}(\mathbf{D})$ can be simultaneously increased
suggests that $R^{\rm tw}(\mathbf{B}, \mathbf{D})$ can be
increased. This contradicts the fact that $\mathbf{D}^*$ is the
optimal solution that maximizes $R^{\rm tw}(\mathbf{B},
\mathbf{D})$ with $\mathbf{D}^*$ subject to the related
constraints. Therefore, $\mathbf{D}^*$ must maximize
$\min\limits_l \{1/\mu_l\}$ subject to \eqref{e:ProfTh3cons}.

Proof of property 4. It can be seen that the maximum achievable
$1/\mu_j$ subject to the constraints
%\begin{subequations}
%\begin{align}
\begin{equation}\label{e:Th3P4c}
R^{\rm ma}(\mathbf{D})\!\geq \! R^{\rm obj}, \quad
\text{Tr}(\mathbf{D}_l)\!\leq \! P_l^\mathrm{max}, \forall l
\end{equation}
%\end{align}
%\end{subequations}
is a non-increasing function of $R^{\rm obj}$. If
$1/\mu_i^*\leq1/\mu_j^*$, according to property 1 of this theorem
and the fact that $1/\mu_\mathrm{ma} <\max_l\limits \{ 1/\mu_l\}$,
it can be shown that $1/\mu_j^*>1/\mu_\mathrm{ma}^*$. Since
$1/\mu_\mathrm{ma}^0 >1/\mu_j^0$ and the maximum achievable
$1/\mu_j$ is a non-increasing function of $R^{\rm obj}$, there
exists $\tilde{\mathbf{D}}$ such that $1/ \mu_j^* \geq1/
\tilde{\mu}_j$ and $1/\tilde{\mu}_j=1/\tilde{\mu}_\mathrm{ma} \geq
1/ \mu_\mathrm{ma}^*$. Using $1/\mu_\mathrm{ma}
<\max\limits_l\{1/\mu_l\}$ from Lemma~1 in Part~I \cite{PartI}, it can be shown that
$1/\tilde{\mu}_i>1/\tilde{\mu}_j=1/\tilde{\mu}_\mathrm{ma}$ at
this point. Since the maximum $\bar{R}_{j\mathrm{r}}
(\mathbf{D}_j)$ cannot achieve $\hat{R}_{\mathrm{r}i}(\lambda^0)$
in the problem \eqref{e:Sub3}, it can be seen that
$1/\tilde{\mu}_j=1/\tilde{\mu}_\mathrm{ma} <1/\lambda^0$. In such
a case, the optimal strategy of the relay is to use
$1/\lambda_l=1/\tilde{\mu}_\mathrm{ma}<1/\lambda^0, \forall l$,
which does not consume the full power of the relay. Therefore,
according to property 2 of this theorem, the $R^{\rm
tw}(\mathbf{B}, \mathbf{D})$ that can be achieved, specifically
$R^{\rm ma}(\tilde{\mathbf{D}})$, in the case that
$1/\tilde{\mu}_j=1/\tilde{\mu}_\mathrm{ma}$ is not the maximum
that $R^{\rm tw}(\mathbf{B}, \mathbf{D})$ can achieve. Moreover,
since $1/\tilde{\mu}_\mathrm{ma}\geq 1/\mu_\mathrm{ma}^*$, it can
be seen that $R^{\rm ma}(\mathbf{D}^*)\leq R^{\rm ma}
(\tilde{\mathbf{D}})$. As a result, $R^{\rm tw}(\mathbf{B}^*,
\mathbf{D}^*)=R^{\rm ma}(\mathbf{D}^*)\leq R^{\rm ma}
(\tilde{\mathbf{D}})$. Using the above-proved fact that $R^{\rm
ma} (\tilde{\mathbf{D}})$ is not the maximum that $R^{\rm tw}
(\mathbf{B}, \mathbf{D})$ can achieve, this result obtained under
the assumption $1/\mu_i^*\leq1/\mu_j^*$ contradicts the assumption
that $\mathbf{B}^*$ and $\mathbf{D}^*$ are optimal. Therefore, the
assumption that $1/\mu_i^*\leq1/\mu_j^*$ must be invalid.
\hfill$\blacksquare$

\vspace{-1mm}
\subsection{Proof of Theorem~3}\label{s:PTh3}
The proof follows the same route as the proof of Theorem~2.

Proof of property 1. As there exists no $\lambda_j$ which
satisfies the constraints in \eqref{e:SubII3con}, it can be seen
that $\sum\limits_l\hat{R}_{\mathrm{r}l}(\lambda_l)$ cannot
achieve $R^{\rm ma}(\mathbf{D}^0)$ subject to the constraint
$\lambda_i=\mu_j^0$, which is necessary as stated in Lemma~2.
Therefore, it is necessary that $1/\mu_\mathrm{ma}^* < 1/
\mu_\mathrm{ma}^0$. Given that $1/ \mu_\mathrm{ma}^* < 1/
\mu_\mathrm{ma}^0$, it can be shown that the resulting $R^{\rm
tw}(\mathbf{B}, \mathbf{D})$ is not maximized if $1/
\mu_\mathrm{ma}^* \leq \min\limits_{l}\{1/\mu_l^*\}$. Therefore,
it is necessary that $1/\mu_\mathrm{ma}^* > \min\limits_{l}
\{1/\mu_l^*\}$.

Proof of properties 2-3 from Section~\ref{s:PTh2} can be applied
here after we substitute all $\lambda^0$ therein to
$\mu_\mathrm{ma}^0$. Proof of property 4 of Theorem~2 can be
directly applied here. Thus, property 2 of Theorem~3 is proved.
\hfill$\blacksquare$


\begin{thebibliography}{1}

\bibitem{TWR_PRT}
B.~Rankov and A.~Wittneben, ``Spectral efficient protocols for
half-duplex fading relay channels,'' {\it IEEE J. Sel. Areas
Commun.}, vol.~25, no.~2, pp.~379--389, Feb.~2007.


\bibitem{PartI}
J.~Gao, S.~A.~Vorobyov, H.~Jiang, J.~Zhang, and M.~Haardt, ``Sum-Rate maximization for MIMO DF two-way relaying with minimum power consumption: Part I - Relay Optimization''.


\bibitem{DFTWR_BC}
\textcolor{black}{T.~J. Oechtering, R.~F. Wyrembelski, and H.~Boche, ``Multiantenna bidirectional broadcast channels -- Optimal transmit strategies,'' {\it IEEE. Trans. Signal Process.}, vol.~57, no.~5, pp.~1948-1958, May~2009.}


\bibitem{SRJoint1}
S.~Xu and Y.~Hua, ``Optimal design of spatial source-and-relay matrices for a non-regenerative two-way MIMO relay system,'' {\it IEEE Trans. Wireless Commun.}, vol.~10, no.~5, pp.~1645-1655, May 2011.

\bibitem{SRJoint2}
C.~Y.~Leow, Z.~Ding, and K.~K.~Leung, ``Joint beamforming and
power management for nonregenerative MIMO two-way relaying
channels,'' {\it IEEE Trans. Veh. Technol.}, vol.~60, no.~9,
pp.~4374--4383, Nov.~2011.

\bibitem{SRJoint3}
R.~Wang and M.~Tao, ``Joint source and relay precoding designs for
MIMO two-way relaying based on MSE criterion,'' {\it IEEE. Trans.
Signal Process.}, vol.~60, no.~3, pp.~1352--1365, Mar.~2012.
%

\bibitem{SRJoint4}
J.~Zou, H.~Luo, M.~Tao, and R.~Wang, ``Joint source and relay optimization for non-regenerative MIMO two-way relay systems with imperfect CSI,''  {\it IEEE Trans. Wireless Commun.}, accetped.


\bibitem{ProtComp}
S. J. Kim, N. Devroye, P. Mitran, and V. Tarokh, ``Achievable Rate Regions and Performance Comparison of Half Duplex Bi-Directional Relaying Protocols,'' {\it IEEE Trans. Inf. Theory}, vol. 57, no. 10, pp. 6405-6418, Oct. 2011.


\bibitem{Jie2012}
J.~Gao, J.~Zhang, S.~A.~Vorobyov, H.~Jiang, and M.~Haardt, ``Power
allocation/beamforming for decode-and-forward MIMO two-way
relaying: Relay optimization and network optimization,'' {\it IEEE
Global Telecommunications Conf.}, Anaheim, CA, USA, Dec.~2012,
accepted, available at http://www.ece.ualberta.ca/{\texttildelow}vorobyov/GLOBECOM12.pdf.
%

\bibitem{Mimo_Cap}
A.~Goldsmith, S.~A.~Jafar, N.~Jindal, and S.~Vishwanath,
``Capacity limits of MIMO channels,'' {\it IEEE J. Sel. Areas
Commun.}, vol.~21, no.~5, pp.~684--702, June~2003.

\bibitem{XORCmp1}
\textcolor{black}{M. Chen and A. Yener, ``Power allocation for F/TDMA multiuser two-way relay networks,'' {\it IEEE Trans. Wireless
Commun.}, vol.~9, no.~2, pp.~546-551, Feb. 2010.}

\bibitem{XORCmp2}
\textcolor{black}{C.-~H. Liu and F. Xue, ``Network coding for two-way relaying: rate region, sum rate and opportunistic scheduling,'' in {\it Proc. IEEE Int. Conf. Commun. 2008}, Beijing, China, May 2008, pp.1044-1049.}

\bibitem{XORCmp3}
\textcolor{black}{J. Liu, M. Tao, Y. Xu, and X. Wang, ``Superimposed XOR: a new physical layer network coding scheme for two-way relay channels,'' in {\it Proc. Global Telecommun. Conf. 2011}, Honolulu, USA, Dec. 2009.}

%\bibitem{XORCmp4}
%R. Wang,  M. Tao, and Y. Liu, ``Optimal linear transceiver designs for cognitive two-way relay networks,'' {\it IEEE. Trans.
%Signal Process.}, http://ieeexplore.ieee.org/stamp/stamp.jsp?tp=\&arnumber=6340357.

\bibitem{WF_MAC}
W.~Yu, R.~Wonjong, S.~Boyd, and J.~M.~Cioffi, ``Iterative
water-filling for Gaussian vector multiple-access channels,'' {\it
IEEE Trans. Inf. Theory}, vol.~50, no.~1, pp.~145-152, Jan.~2004.
\end{thebibliography}
\end{document}